\documentclass[aps,twocolumn,showpacs,amsmath,amssymb,superscriptaddress,longbibliography,prl]{revtex4-1}
\usepackage[colorlinks=true,citecolor=blue,linkcolor=blue,hypertexnames=false]{hyperref}
\usepackage{amsmath}
\usepackage{dsfont}
\usepackage{hyperref}
\usepackage{textcomp}
\usepackage{tabularx}
\usepackage{graphicx}
\usepackage{soul}
\usepackage{feynmp-auto}
\usepackage[compat=1.0.0]{tikz-feynman}
\usepackage{footmisc}
\usepackage[export]{adjustbox}
\usepackage{wrapfig}

\newcommand{\bk}{{\bf k}}
\newcommand{\bx}{{\bf x}}
\newcommand{\bq}{{\bf q}}

\newcommand{\bQ}{{\bf Q}}

\newcommand{\bp}{{\bf p}}

\newcommand{\ve}{{\varepsilon}}

\def \be{\begin{equation}}
\def \ee{\end{equation}}
\def \bea{\begin{eqnarray}}
\def \eea{\end{eqnarray}}

\newcommand{\old}[1]{\iffalse #1 \fi}
\begin{document}

\newcommand\rmf[1]{\textcolor{magenta}{#1}}

\title{Instability of the critical Ngai's coupling and two-boson mechanism in metals}

\author{Phu Nguyen}
\affiliation{Department of Physics, Carnegie Mellon University, Pittsburgh, Pennsylvania 15213, USA}

\author{Vladyslav Kozii}
\affiliation{Department of Physics, Carnegie Mellon University, Pittsburgh, Pennsylvania 15213, USA}

\date{\today}

\begin{abstract}
We study the properties of a Fermi liquid coupled to a quantum critical boson via the two-boson interaction known as Ngai's coupling. We find that the original quantum critical point is generally unstable, resulting in a finite-momentum spatially modulated state unless two conditions are satisfied: (i) the critical boson is polar and transverse, and (ii) the ratio of the Fermi velocity to the transverse-boson velocity is sufficiently large. If these conditions hold and the uniform state remains stable, we demonstrate that the system enters a strong-coupling regime below a certain energy scale. In this regime, we discuss a self-consistent solution at criticality in two-dimensional systems and show that the critical boson field develops a nontrivial anomalous dimension, $\eta=1/2$. Our findings highlight the significant role of two-boson coupling in critical theories, challenging the conventional view that its effects are subdominant to linear coupling.
\end{abstract}

\maketitle
{\it Introduction.---} The impact of two-boson-mediated electron-electron interactions on transport, optical properties, and superconductivity in a Fermi liquid has been the subject of intensive research in recent years. This interaction has been studied in various systems, primarily in the context of electron-phonon coupling~\cite{Ngai74,EntinWohlman1983,Entin-Wohlman_1985,Kuklov1989,GogolinPolaron91,GogolinMass91,Adolphs_2013,AdolphsBerciu2014,AdolphsBerciu20142,Nazaryan2021,KiselovFeigelman21,MaslovNgai2021,VolkovColeman22,han2024quantum,Svistunov2023}. Unlike the conventional linear electron-boson interaction, which is often limited or entirely forbidden by symmetry, the two-boson coupling is always present, regardless of the boson's nature or the system's symmetry. This coupling is linear in electron density but {\it quadratic} in boson fields. It has attracted particular interest in the low-density superconductor SrTiO$_3$, where it has been proposed that Cooper pairing is mediated by the exchange of two transverse polar optical phonons, which soften near the ferroelectric quantum critical point (QCP). In this context, the mechanism is often referred to as {\it Ngai's coupling}~\cite{Ngai74}, which motivates our study. It has been shown that, near the critical point, such two-boson processes can mediate long-range attraction between electrons, potentially compensating for their low density~\cite{KiselovFeigelman21,VolkovColeman22}.


In this paper, we investigate the properties of a metallic system coupled to massless bosons at a QCP via the two-boson interaction. A key strength of our analysis lies in its broad applicability to various types of critical points (e.g., magnetic, nematic, ferroelectric) since it does not depend on the specific nature of the critical boson or impose symmetry constraints. We also examine the case of a polar transverse boson, such as the transverse optical phonon near the ferroelectric instability in a polar crystal. While previous studies have treated these phonons within a mean-field framework~\cite{MaslovNgai2021,KiselovFeigelman21,VolkovColeman22}, we demonstrate that quantum critical fluctuations play a crucial role at low energies and must be carefully accounted for.



Our main finding is that the two-boson coupling to the electron density renders the original critical point unstable, leading to a nonuniform state in which the order parameter develops a spatial modulation at a finite wave vector. To justify our analytical calculations, we focus on the limit of weak bare coupling. We show that in $d=3$ (spatial) dimensions, the ordering wave vector has an exponential dependence on the coupling constant, whereas in $d=2$, this dependence follows a power law. We also find that the uniform  state (with zero wave vector) can be stabilized in the case of a polar transverse boson, provided that the ratio of the Fermi velocity to the critical-boson velocity exceeds a certain threshold. Since this parameter can be experimentally controlled, our theory makes a potentially testable prediction. At the uniform QCP, we demonstrate that the dominant vertex corrections enhance the coupling constant, driving the system into the strong-coupling regime at low energies. These results apply to both $d=2$ and $d=3$. In $d=2$, we also derive a self-consistent low-energy solution for the boson and fermion propagators, showing that the boson field acquires an anomalous critical dimension $\eta = 1/2$.


{\it Model description.---} We start our analysis by formulating an effective low-energy field theory in $d$ spatial dimensions. The imaginary-time (Matsubara) action at temperature $T=0$ takes form

\begin{align}
&S = S_\psi + S_\phi + S_{\psi-\phi}, \label{Eq:action} \\
&S_\psi = \sum_{n=1}^N\int \frac{d\omega d^d \bk}{(2\pi)^{d+1}} \psi^\dagger_n(\omega,\bk) \left(i\omega - \xi_k  \right)\psi_n(\omega,\bk), \nonumber \\ &S_\phi = \frac12\int \frac{d\Omega d^d \bQ}{(2\pi)^{d+1}} | \boldsymbol{\phi}(\Omega,\bQ)|^2 \left(r + \Omega^2 + v_B^2 Q^2  \right), \nonumber \\ &S_{\psi-\phi} = \frac{\lambda}2 \sum_{n=1}^N \int d\tau d^d\bx \,\psi^\dagger_n(\tau,\bx) \psi_n(\tau,\bx) \boldsymbol{\phi}^2(\tau,\bx). \nonumber
\end{align}
Grassmann fields $\psi^\dagger_n$ and $\psi_n$, with $n=1,\ldots,N$, describe $N$ identical flavors of itinerant (metallic) electrons. We introduced $\xi_k = \ve_k-\ve_F \approx v_F (k-k_F),$ where $\ve_k$ is the single-electron spectrum (which we assume to be isotropic) and $v_F$, $k_F$ and $\ve_F$ are Fermi velocity, momentum and energy, respectively. The real fields $\phi_i(\tau,\bx)$ describe bosonic degrees of freedom, the parameter $r$ measures the bare distance to the criticality (square of the gap in the bare bosonic spectrum) and $v_B$ is the bare bosonic velocity at criticality.  We also use a shorthand notation $\boldsymbol{\phi}^2 = \sum_{i=1}^M \phi_i^2$, where $M$ is the number of bosonic components. This field may represent, e.g., magnetization near ferromagnetic QCP or nematic order parameter near nematic QCP. We use the units with $\hbar=1$ throughout this work.

We additionally study the scenario where the bosonic field is {\it polar} (vector-type) and {\it transverse} with respect to its momentum. In this case, the action is formally given by the same expression as in Eq.~\eqref{Eq:action}, but now the fields $\phi_i$ should be replaced with the transverse fields $\phi_i^T$:
\be  
\phi_i^T(\Omega,\bQ) = P^T_{ij}(\bQ) \phi_j(\Omega, \bQ), \label{Eq:phi} \tag{\theequation a}
\ee 
where $P^T_{ij}(\bQ) = \delta_{ij} - \hat Q_i \hat Q_j$ is the projector onto transverse modes, $\hat \bQ = \bQ/|\bQ|$ is the unit vector, and the summation over repeated indices is implied. This expression only makes sense if the number of bosonic components is equal to the number of spatial dimensions $M = d$. In the context of a ferroelectric QCP, the field ${\boldsymbol \phi}$ is proportional to the optical phonon displacement, and only the transverse modes become gapless at the critical point, while the longitudinal mode remains massive due to splitting between the longitudinal and transverse optical phonons~\cite{AMbook}.

We also neglect the possibility of superconductivity, assuming that it is suppressed by, e.g., magnetic field. The pairing problem and interplay between the two-boson criticality and superconductivity will be considered in a separate work. 

\begin{figure}
 \begin{center}
   \includegraphics[width=1\linewidth]{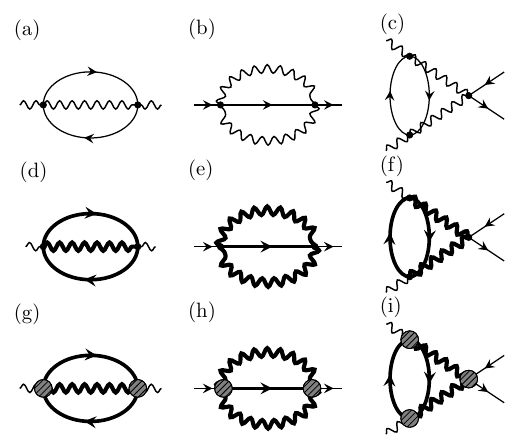}
 \end{center}
\caption{The lowest-order nontrivial diagrams demonstrating fermion self-energy, boson self-energy, and the leading vertex correction. Diagrams (a)-(c) contain bare fermion (thin straight line) and boson (thin wavy line) propagators and bare vertices only. Diagrams (d)-(f) contain dressed (self-consistent) propagators (bold lines) and bare vertices. These are used to derive self-consistent solution~\eqref{Eq:D2d} for a transverse boson in $d=2$. Diagrams (g)-(i) contain dressed propagators and  vertices (gray shaded dots). Fermion self-energy is regular and preserves Fermi liquid in all  relevant cases  (except for diagram (b) in $d=2$), so we do not distinguish between bare and dressed fermion propagators in our calculations.}
 \label{Fig:diagrams}
\end{figure}

{\it Instability and strong-coupling regime in $d=3$.---} First, we calculate the lowest-order nontrivial contribution to (bare) boson self-energy shown in Fig.~\ref{Fig:diagrams}(a). Using Eq.~\eqref{Eq:action} at the critical point, $r=0$, we find with logarithmic accuracy that
\begin{align}  
&\Pi (i\Omega, \bQ) \approx \Pi(0,0) \label{Eq:Pi3D} \\ &+\frac{\lambda^2 N N_0 v_F}{24 \pi^2 v_B^2 (v_F + v_B)^2} \left(3 \Omega^2 \ln \frac{v_B \Lambda}{|\Omega|} -  v_B^2 Q^2 \ln \frac{\Lambda}Q \right).\nonumber 
\end{align}
Here $N_0$ is the density of states at the Fermi level per one fermion flavor and $Q = |\bQ|$. We delegate the details of the calculations to the Supplemental Materials (SM)~\cite{SM}. In our convention, the dressed (fully renormalized) boson propagator $D$ is related to its self-energy $\Pi$ as $D^{-1} =D_0^{-1} + \Pi$, where the bare propagator at criticality equals $D_0^{-1}(i\Omega, \bQ) = \Omega^2 + v_B^2 Q^2$. The constant term $\Pi(0,0)\propto \lambda^2 \Lambda^2$ only shifts the position of the critical point. We introduced the ultraviolet momentum cutoff $\Lambda$ and assumed that $\Lambda \ll k_F$ (in addition to $Q, |\Omega|/v_B \ll \Lambda$) to simplify the calculations. This scale restricts the momentum range where the bosonic dispersion can be approximated as linear, according to Eq.~\eqref{Eq:action}.

Equation~\eqref{Eq:Pi3D} in the static limit $\Omega = 0$ indicates that the original critical point at $Q=0$ is preempted by the finite-momentum one with $Q=Q_0$, where 
\be 
Q_0 \sim \Lambda \exp\left( - \frac{24 \pi^2 v_B^2(v_F + v_B)^2}{\lambda^2 N N_0 v_F}  \right). \label{Eq:Qcr}
\ee
Our analytical approach is only meaningful if the exponential factor is small, $\lambda^2 N N_0 v_F \ll 24 \pi^2 v_B^2 (v_F + v_B)^2$, implying that $Q_0 \ll \Lambda$. Otherwise, the effective low-energy field-theory formulation is not applicable. 

Next, we demonstrate that the result becomes qualitatively different if the bosonic fields are transverse, i.e., given by Eq.~\eqref{Eq:phi}. From a technical perspective, it means that the boson self-energy, as well as the bare boson propagator entering it, must be projected onto the transverse sector by $P_{ij}^T$. The result reads as~\cite{SM}
\begin{align}  
&\Pi_T (i\Omega, \bQ) \approx \Pi_T(0,0) \label{Eq:Pit2D} \\ &+ \alpha \left[g^T_{\Omega}\left(\frac{v_F}{v_B}  \right) \Omega^2 \ln \frac{v_B \Lambda}{|\Omega|} + v_B^2 g^T_Q\left(\frac{v_F}{v_B}  \right) Q^2 \ln \frac{\Lambda}Q\right],\nonumber 
\end{align}
and index $T$ stands for ``transverse''. Here we introduced 
\begin{align}  
&\alpha = \frac{\lambda^2 N N_0}{12 \pi^2 v_B^3}, \qquad g^T_{\Omega}(x) = \frac{x}{(x+1)^2}, \label{Eq:alpha} \\ &g^T_Q(x) = \frac15 \left( \frac{\ln(1+x)}x-\frac{1+3x}{(1+x)^2}  \right). \nonumber
\end{align}

The function $g^T_Q(x)$ changes sign at $x^* \approx 13.3$. We note that at $v_F/v_B < x^*$ the system is unstable again toward the finite-momentum state with the ordering wave vector 
\be 
Q^T_0 \sim \Lambda \exp\left( - \frac{1}{\alpha |g^T_Q(v_F/v_B)|}  \right). \label{Eq:QTcr}
\ee
In the context of a ferroelectric QCP, this nonuniform state has previously been considered to originate from Rashba-type linear coupling and dubbed {\it ferroelectric density wave}~\cite{KoziiKleinFernandesRuhman22,KleinKoziiRuhmanFernandes23}. 

In the case of a larger Fermi velocity, $v_F/v_B > x^*$, the uniform (zero-momentum) state remains stable, at least within our lowest-order perturbative analysis. However, below exponentially small momentum and frequency of the order $Q^T_0$, the perturbative corrections become comparable to the bare propagator. Consequently, the summation of the leading-order logarithmical terms is desirable. To achieve this goal, we first calculate the lowest-order perturbative corrections to the fermion self-energy and the interaction vertex shown in Figs.~\ref{Fig:diagrams}(b)-(c).

We find that the vertex correction $\delta \lambda$ contains a large logarithmic factor:
\be  
\frac{\delta\lambda}{\lambda} \approx  \alpha g^T_\lambda\left(\frac{v_F}{v_B} \right)\ln \frac{\Lambda}{\sqrt{Q^2 + \Omega^2/v_B^2}}, \,\, g^T_\lambda(x) = \frac{x}{x+1}. \label{Eq:lambda}
\ee 
In contrast, the fermion self-energy is regular and small as long as the bare coupling is weak. It just insignificantly renormalizes fermionic parameters while preserving the Fermi-liquid behavior. 

Focusing on logarithmic terms only, we are able to derive the effective renormalization group equations in the limit $\alpha \ll 1$. This approach is equivalent to summing up ``parquet'' (without intersections) diagrams~\cite{Note1}  or deriving self-consistent equations for the renormalized (dressed) self-energies and vertex correction shown in Figs.~\ref{Fig:diagrams}(g)-(i)~\cite{Abrikosov1971,Hosur2012}. Solving these equations approximately, we find that the transverse boson propagator and interaction vertex renormalize as~\cite{SM} 
\begin{align}
D_T^{-1}(i\Omega,Q) &\to \frac{\Omega^2}{(1- \alpha \hat g l)^{g_\Omega^T/\hat g}} + \frac{v_B^2Q^2}{(1- \alpha \hat g l)^{g_Q^T/\hat g}}, \nonumber \\ \lambda &\to \frac{\lambda}{(1- \alpha \hat g l)^{g_\lambda^T/\hat g}},
\end{align}
where $l = \ln(\Lambda/\sqrt{Q^2 + \Omega^2/v_B^2})$ and $\hat g = 2 g^T_\lambda - g^T_\Omega/2 - 3g^T_Q/2$. Different $g$-functions are given by Eqs.~\eqref{Eq:alpha} and~\eqref{Eq:lambda}, with the argument $v_F/v_B$. 

Despite the absence of any indication for the finite-momentum instability at $v_F/v_B > x^*$, our analysis shows that the system enters the strong-coupling regime at the scale where $\alpha \hat g l \sim 1$. Up to a numerical factor in front of $1/\alpha$, it exactly coincides with the exponential momentum scales set by Eqs.~\eqref{Eq:Qcr} and~\eqref{Eq:QTcr}.

{\it Instability and self-consistent solution in $d=2$.---} The main results for two-dimensional systems are qualitatively similar to the results obtained in three dimensions. However, the important exponential scales~\eqref{Eq:Qcr} and~\eqref{Eq:QTcr} are replaced by the power-law dependence on the coupling constant because of stronger quantum fluctuations. 

The lowest-order nontrivial contribution to boson self-energy is shown in Fig.~\ref{Fig:diagrams}(a) and equals
\begin{align}  
&\Pi(i\Omega,\bQ) \approx \Pi(0,0) \label{Eq:Pi2D}\\ &+ \frac{\lambda^2 N N_0}{v_B^2}\left[f_\Omega\left( \frac{v_F}{v_B}\right) |\Omega| + f_Q\left( \frac{v_F}{v_B}\right)v_B Q\right].   \nonumber
\end{align}
Explicit analytic expressions for $f_Q(x)$ and $f_\Omega(x)$ are presented in SM~\cite{SM}, but we note here that $f_Q(x) < 0$ for all $x$. This lowest-order perturbative analysis indicates, again, toward a (static) finite-momentum instability with the ordering wave vector 
\be  
Q_0 = \frac{\lambda^2 N N_0}{2v_B^3} \left|f_Q\left( \frac{v_F}{v_B} \right)\right|. \label{Eq:Qcr2D}
\ee
Constant term $\Pi(0,0)\propto \lambda^2 \Lambda$ merely implies a small shift of the bare critical point.

Similarly to the case $d=3$, the results are more diverse if the boson is transverse and given by Eq.~\eqref{Eq:phi}. The lowest-order boson self-energy in this case equals
\begin{align}  
&\Pi_T(i\Omega,\bQ) \approx \Pi_T(0,0) \\ &+ \frac{\lambda^2 N N_0}{v_B^2}\left[ f^T_\Omega\left( \frac{v_F}{v_B}\right) |\Omega|  + f^T_Q\left( \frac{v_F}{v_B}\right)v_B Q \right],   \nonumber
\end{align}
where the function $f^T_Q(x)$ changes sign from negative to positive at $x^* \approx 9.43$. This perturbative analysis indicates that the uniform QCP might remain stable if the ratio $v_F/v_B$ is sufficiently large. 

Unlike the case $d=3$, where the boson self-energy and vertex corrections are only logarithmic, the perturbation theory completely breaks down in $d=2$, even if the state remains uniform. We see that at sufficiently small frequencies/momenta, below the scale given parametrically by Eq.~\eqref{Eq:Qcr2D}, power-law perturbative correction completely dominates over the bare bosonic propagator. Furthermore, one can easily see from the dimensional analysis that the $\Omega,Q$-linear expression for $\Pi_T$ (and consequently, $D_T^{-1}$) is {\it not} a self-consistent solution when plugged back into the diagram in Fig.~\ref{Fig:diagrams}(a) (or~\ref{Fig:diagrams}(d)).

 We next look for a self-consistent low-energy solution for a transverse boson at the uniform QCP, neglecting vertex corrections for now. Simple power counting dictates the scaling $\Pi_T \propto \Omega^{3/2},Q^{3/2}$ at low frequencies and momenta. We use the following ansatz for analytical convenience: 
 \be  
D_T^{-1}(i\Omega,\bQ) \approx \delta\Pi_T(i\Omega, \bQ) = \Omega_0^{1/2} (\Omega^2 + c^2 Q^2)^{3/4}, \label{Eq:D2d}
 \ee 
 where $\delta\Pi_T(i\Omega, \bQ) \equiv \Pi_T(i\Omega, \bQ) - \Pi_T(0,0)$, and parameters $\Omega_0$ and $c$ (the latter representing the physical velocity of the critical boson) must be found self-consistently. Plugging this expression into the diagram in Fig.~\ref{Fig:diagrams}(d) and calculating the frequency and momentum dependence of $\Pi_T$ with the self-consistent propagator $D_T$, we find indeed the solution~\cite{SM}:
\be  
\Omega_0 \approx 0.82\frac{\lambda^2 N N_0}{v_F^2 } , \qquad c \approx 0.072  v_F.  \label{Eq:Omega0c}
\ee 
We stress that it is crucial to have a {\it transverse} boson for this result to hold. Otherwise, the sign in front of $\delta\Pi(0,\bQ)$ is negative, again indicating the finite-momentum instability. 

Equation~\eqref{Eq:D2d} implies that the transverse boson acquires an anomalous dimension $\eta = 1/2$ at the uniform QCP.
This low-energy solution is valid as long as the bare propagator can be neglected compared to the self-energy, $D_0^{-1} \ll \delta\Pi_T$. This condition is satisfied for small frequencies and momenta below $\Omega_0$,
\be  
\max\{|\Omega|, cQ \} \lesssim \Omega_0,
\ee 
which, up to numerical prefactors or velocity ratios, parametrically coincide with the scale from Eq.~\eqref{Eq:Qcr2D}. At higher frequencies and momenta, the dressed boson propagator coincides with the bare one. 

The fermion self-energy with the self-consistent boson propagator~\eqref{Eq:D2d} is given by the diagram shown in Fig.~\ref{Fig:diagrams}(e). We find that it is regular and small below $\Omega_0$ at weak coupling, thus preserving the Fermi-liquid behavior and only renormalizing bare fermionic parameters. At energies larger than $\Omega_0$, the self-energy acquires logarithmic corrections~\cite{SM}.

Finally, we discuss the modification of our self-consistent solution at the uniform QCP in $d=2$ due to {\it vertex corrections}. The leading lowest-order correction, with the self-consistent boson propagator~\eqref{Eq:D2d}, is shown in Fig.~\ref{Fig:diagrams}(f) and, for $\sqrt{\Omega^2 + c^2 Q^2} \lesssim \Omega_0$, equals
\be 
\delta \lambda (i\Omega,\bQ) \approx 5.57 \lambda \ln\frac{\Omega_0}{\sqrt{\Omega^2 + c^2 Q^2}}, \label{Eq:deltalambda2d}
\ee 
where $\Omega$ and $Q$ are frequency and momentum transferred between the two bosonic (or two fermionic) external lines.

We see that the first perturbative correction around the self-consistent solution is already logarithmically larger than the bare vertex, indicating that the system enters the {\it strong-coupling regime} at the scale $\Omega_0$. The large-$N$ assumption in this theory does not allow us to neglect vertex corrections since they also have fermionic loops. Furthermore, due to the self-consistency condition~\eqref{Eq:Omega0c}, $\delta \lambda$ does not contain any extra powers of $\lambda$. 

We attempt to account for logarithmic vertex corrections by substituting bare vertices in diagrams~\ref{Fig:diagrams}(d) and~\ref{Fig:diagrams}(f) with fully renormalized ones, as shown in  diagrams~\ref{Fig:diagrams}(g) and~\ref{Fig:diagrams}(i). We only consider ``parquet'' diagrams without intersections and with the maximal number of fermionic loops, as only these contribute large logarithmic factors. We find, indeed, that the low-energy self-consistency equations are satisfied if we modify the boson propagator and the coupling constant as~\cite{SM}
\begin{align}
&D_T(i\Omega,\bQ) \to \frac{\ln\frac{\Omega_0}{\sqrt{\Omega^2 + c^2 Q^2}}}{\Omega_0^{1/2} (\Omega^2 + c^2 Q^2)^{3/4}}, \nonumber \\ &\lambda \to \Gamma(i\Omega,\bQ) \approx \frac{\tilde \lambda}{\ln\frac{\Omega_0}{\sqrt{\Omega^2 + c^2 Q^2}}}. \label{Eq:scsolution}
\end{align}
Here we define $\tilde \lambda \approx -\lambda/5.57$, and $\Omega_0$ should be redefined by replacing $\lambda^2$ with $\tilde \lambda^2$ in Eq.~\eqref{Eq:Omega0c}. 
The fermion self-energy  remains regular and small. 

Although our equations are formally consistent with the analytical solution~\eqref{Eq:scsolution}  in the low-energy limit, their smooth connection to the high-energy  scale has yet to be determined. Indeed, since correction~\eqref{Eq:deltalambda2d} pushes the system to a strong coupling and $\tilde \lambda$ has a different sign compared to $\lambda$, Eq.~\eqref{Eq:scsolution} should only be considered as a guess rather than a mathematically rigorous solution at the QCP. Addressing this important question seems to be approachable only numerically, and we leave it to a future work. 

{\it Conclusion.---} 
We studied the critical properties of a QCP coupled to the density of electrons with a finite Fermi surface through a two-boson coupling in both $d=2$ and $d=3$ spatial dimensions. Our results are applicable to a large family of QCPs as we do not specify the nature of a critical boson. We found that the original critical point is generically unstable to the formation of a finite-momentum state modulated in space. Additionally, we find that if the boson is polar and transverse to the direction of propagation, the original (uniform) critical point may be stabilized if the ratio between Fermi velocity and bosonic velocity is sufficiently large. In this case, we show that the system flows to the strong-coupling regime at low energies. Our dimensional analysis also shows that, in $d=2$, a critical boson acquires a nontrivial anomalous dimension $\eta = 1/2$.

Our work opens up a plethora of questions for future study. To investigate the properties of the ordered state and the nature of the transition, one has to derive an effective Ginzburg-Landau theory for the bosonic field, which is expected to have non-analyticities after integrating out gapless fermions. Furthermore, to put our analytical study on firmer ground, one can supplement it with the renormalization group analysis, the numerical solution of the self-consistency equations, or even perform a numerical simulation.  Indeed, this theory is a wonderful candidate for Quantum Monte Carlo simulation, as it does not suffer from the sign problem. Finally, it is worth studying how this type of criticality affects or intertwines with superconductivity. This question is especially relevant in doped SrTiO$_3$, where the Ngai's coupling should be compared with other proposed mechanisms for superconductivity~\cite{Fernandesetal2013,BalatskiiPRL2015,RuhmanLee2016,WolfleBalatsky2018,Kedem2018,GastiasoroChubukovFernandes2019,GastiasoroRuhmanFernandesSTO2020}, including those originating from Rashba-like linear coupling~\cite{sahaetal2024,KoziiBiRuhman2019,KoziiFu2015}. Our rough estimates show that the scale from Eqs.~\eqref{Eq:Qcr} or~\eqref{Eq:QTcr} is negligible in SrTiO$_3$, indicating that the mean-field description is possibly sufficient for this material. However, a more careful analysis is required to arrive at a reliable conclusion. Furthermore, our work may have implications for other materials that exhibit superconductivity near a critical point, such as strong ferroelectric tendency in KTaO$_3$ interfaces~\cite{KTO2021,KTOLevy2022} or nematicity in FeSe~\cite{ColdeaFeSe2017}.


\begin{acknowledgements}
The authors sincerely thank Jonathan Ruhman, Leo Radzihovsky, Max Metlitskii, Grigory Tarnopolskiy and  Ira Rothstein for valuable discussions and comments on the manuscript. 
\end{acknowledgements}

\bibliography{Ngai}

\newpage

\begin{widetext}

\begin{center}
\textbf{\large Supplemental Materials for ``Instability of the critical Ngai coupling and two-boson mechanism in metals''} 
\end{center}
\setcounter{equation}{0}
\setcounter{figure}{0}
\setcounter{table}{0}

\makeatletter
\renewcommand{\theequation}{S\arabic{equation}}
\renewcommand{\thefigure}{S\arabic{figure}}
\renewcommand{\thetable}{S\Roman{table}}


In this Supplemental Materials, we discuss the technical details of the diagrammatic calculations that lead to the results presented in the main text. Our starting point is set by Eq.~\eqref{Eq:action}, from which bare Matsubara fermion and boson propagators are obtained: 
\be  
G_0(i\omega, \bk) = \includegraphics[width=1.4cm,valign=top]{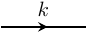} =\frac1{i\omega - \xi_\bk}, \qquad D_0(i\Omega, \bQ) = \includegraphics[width=1.4cm,valign=top]{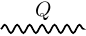} = \frac1{\Omega^2 + v_B^2 Q^2}. \label{SMEq:G0D0}
\ee 
We derive the boson and fermion self-energies, both bare and dressed (self-consistent) where appropriate, as well as the vertex corrections. We consider the cases of both $d=3$ and $d=2$ spatial dimensions. We pay special attention to the case of a transverse boson given by Eq.~\eqref{Eq:phi}, as it offers the most interesting physical implementations. We use the following convention for the Dyson equations:
\be  
D^{-1}(i\Omega, \bQ) = D_0^{-1}(i\Omega, \bQ) + \Pi(i\Omega, \bQ), \qquad G^{-1}(i\omega, \bk) = G^{-1}_0(i\omega, \bk) + \Sigma(i\omega, \bk),
\ee 
where $D$ and $G$ are fully renormalized boson and fermion propagators, correspondingly, while $\Pi$ and $\Sigma$ are boson and fermion self-energies. 

Bosonic frequencies and momenta are restricted by the ultraviolet scale $\Lambda$, i.e., $|\Omega|/v_B, Q< \Lambda$. We also assume that $\Lambda \ll k_F$ for calculational simplicity, where $k_F$ is the Fermi momentum. Throughout this paper, we work in the zero-temperature limit, so the summation over Matsubara frequencies can be replaced by integration.  

\section{I. Calculation in $\bf d=3$ spatial dimensions}
We start our analysis by considering the case of $d=3$ spatial dimensions and an isotropic (in contrast to transverse) bosonic field, as given by Eq.~\eqref{Eq:action} of the main text. The bosonic field generically has $M$ independent components, so the boson propagator and its self-energy are matrices in the component space. However, in case of an isotropic boson, all the correlation functions and vertex corrections are proportional to the Kronecker delta in the component space, so we drop it for brevity. This conclusion does not hold for the transverse boson considered in the following.

\subsection{(a) Boson self-energy}

The lowest-order (one-loop) contribution to the bare boson self-energy $\Pi$ is given by the following diagram: 
\be  
\includegraphics[width=1.2cm,valign=top]{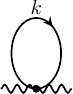} = \lambda N \int \frac{d\ve d^3\bk}{(2\pi)^{4}}  G_0(i\ve,\bk) = \lambda n,
\ee 
where $n$ is the average fermion density. This contribution is trivial, as it merely renormalizes bare boson mass and shifts the position of the critical point. However, this term might be important because it conveniently allows one to estimate the value of the coupling constant $\lambda$ from the experimental measurements.

The lowest-order {\it nontrivial} contribution (two-loop) is given by 
\be  
\Pi(i\Omega,\bQ) = \includegraphics[width=2.5cm,valign=c]{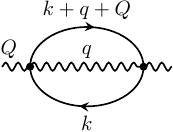} = \lambda^2 N \int \frac{d\omega d^3\bq}{(2\pi)^{4}}  \frac{d\ve d^3\bk}{(2\pi)^{4}} G_0(i\ve,\bk) G_0(i\ve+i\omega + i\Omega, \bk + \bq + \bQ) D_0(i\omega, \bq), 
\ee 
where the factor $1/4$ from the interaction vertices is compensated by the combinatorial symmetry factor (number of different contractions) of the diagram. In the limit $|\omega|/v_F, q \ll k_F$, which is automatically satisfied provided $\Lambda \ll k_F$, we can integrate over $\xi_\bk$ instead of $\bk$ and obtain 
\be  
\int \frac{d\ve d^3\bk}{(2\pi)^{4}} G_0(i\ve,\bk) G_0(i\ve+i\omega, \bk + \bq ) = \includegraphics[width=2.3cm,valign=c]{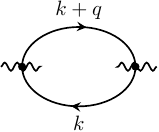} = - N_0\left\{1 - \frac{\omega}{v_F q} \arctan\left(\frac{v_F q}{\omega}\right)\right\}, \label{SMEq:G0G0bubble3D}
\ee 
where $N_0$ is the fermion density of states per one flavor at the Fermi level. Plugging this back and separating out the constant part, $\Pi(i\Omega,\bQ) = \Pi(0,0) + \delta \Pi(i\Omega,\bQ)$, we obtain 
\begin{align}
 \Pi(0, 0) &= -\lambda^2 N N_0 \int \frac{d\omega d^3\bq}{(2\pi)^{4}}  \left\{1 - \frac{\omega}{v_F q} \arctan\left(\frac{v_F q}{\omega} \right)\right\}\frac1{\omega^2 + v_B^2 q^2}, \label{SMEq:PideltaPi3D}\\ \delta\Pi(i\Omega, \bQ) &= -\lambda^2 N N_0 \int \frac{d\omega d^3\bq}{(2\pi)^{4}}  \left\{\frac{\omega}{v_F q} \arctan\left(\frac{v_F q}{\omega}\right) - \frac{\omega+\Omega}{v_F |\bq+\bQ|} \arctan\left(\frac{v_F |\bq+\bQ|}{\omega+ \Omega}\right)\right\}\frac1{\omega^2 + v_B^2 q^2}. \nonumber
\end{align}
To evaluate these integrals, we integrate over $\omega \in (-\infty,\infty)$ first, and then over $q \in (0,\Lambda)$. We find
\be 
\Pi(0,0) = -\frac{\lambda^2 N N_0 \Lambda^2}{8 \pi^2 v_B}\left\{1-\frac{v_B}{v_F}\ln\left(1+\frac{v_F}{v_B}  \right)  \right\}.
\ee 
This term only shifts the position of the critical point. We then calculate the frequency and momentum dependence of the self-energy separately, focusing on the limit $|\Omega|/v_B, Q \ll \Lambda$. We obtain
\begin{align} 
&\delta \Pi(\Omega,0) = \frac{\lambda^2 N N_0 v_F}{8\pi^2 v_B^2 (v_F + v_B)^2} \Omega^2\left\{ \ln\frac{v_B\Lambda}{|\Omega|} + \psi_\Omega\left( \frac{v_F}{v_B} \right)  \right\},  \qquad \psi_{\Omega}(x) = \frac1x + \frac32-\frac{1+2x}{x^2}\ln\left( 1 + x \right), \nonumber \\ &\delta \Pi(0,Q) = -\frac{\lambda^2 N N_0 v_F}{24\pi^2 (v_F + v_B)^2} Q^2\left\{ \ln\frac{\Lambda}Q + \psi_Q\left( \frac{v_F}{v_B} \right)  \right\}, \label{SMEq:Pibare3D}\\ &\psi_Q(x) = \frac1{6(x-1)^2}\left[ 12 x \ln x + (x-1) (-11-x+6(x-1)\ln(x+1))   \right]. \nonumber
\end{align}

Summing up $\Pi(0,0)$, $\delta \Pi(\Omega,0)$ and $\delta \Pi(0,\bQ)$, we reproduce Eq.~\eqref{Eq:Pi3D} of the main text with logarithmic accuracy. We check numerically that this is a good approximation. The sublogarithmic terms in this expression may depend on the regularization scheme. 

Minimizing $D^{-1}(0,Q)$ with respect to $Q$ in the static limit $\Omega=0$, we find instability toward the finite-momentum state with the ordering wave vector 
\be  
Q_0 = \Lambda \exp\left\{ - \frac{24 \pi^2 v_B^2(v_F + v_B)^2}{\lambda^2 N N_0 v_F}  -\frac12 + \psi_Q\left(\frac{v_F}{v_B}\right)\right\}.
\ee

\subsection{(b) Fermion self-energy}
Next, we calculate the bare fermion self-energy and demonstrate that it preserves the Fermi liquid form. The one-loop diagram does not depend on external frequency or momentum and merely renormalizes chemical potential. The lowest-order nontrivial diagram has two loops and is given by
\be  
\Sigma(i\ve,\bk) = \includegraphics[width=2.5cm,valign=c]{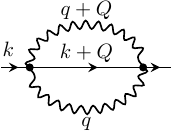} = - \frac{\lambda^2 M}2 \int \frac{d\omega d^3\bq}{(2\pi)^{4}}  \frac{d\Omega d^3\bQ}{(2\pi)^{4}}  D_0(i\omega, \bq)  D_0(i\omega + i\Omega, \bq + \bQ)G_0(i\ve+ i\Omega, \bk + \bQ),
\ee 
where $M$ is the number of bosonic components, and one factor $1/2$ from the vertex is compensated by the symmetry factor of the diagram. As an intermediate step, we find with logarithmic accuracy (e.g., using Feynman parameters)
\be  
\int \frac{d\omega d^3\bq}{(2\pi)^{4}} D_0(i\omega, \bq)  D_0(i\omega + i\Omega, \bq + \bQ) = \includegraphics[width=2.3cm,valign=c]{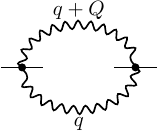} = \frac1{8\pi^2 v_B^3}\ln \frac{v_B \Lambda}{\sqrt{\Omega^2 + v_B^2 Q^2}}, \label{SMEq:bosonicbubble3D}
\ee 
which is valid provided $\sqrt{(\Omega/v_B)^2 +  Q^2} \ll \Lambda$. Then, assuming that $|\bk| \approx k_F$ and performing angular integration, we obtain
\be  
\Sigma(i\ve,\xi_\bk) \approx - \frac{\lambda^2 M}{2^7 \pi^5 v_B^3 v_F}\int_0^{\Lambda} Q dQ \int_{-v_B \Lambda}^{v_B \Lambda}d\Omega \ln\frac{\xi_\bk - v_F Q -i(\ve + \Omega)}{\xi_\bk + v_F Q -i(\ve + \Omega)} \ln \frac{v_B\Lambda}{\sqrt{\Omega^2 + v_B^2 Q^2}}. \label{SMEq:Sigma3Dgen}
\ee 
This expression has two main contributions. The first one comes from large momenta and frequencies $Q \sim |\Omega|/v_B \sim \Lambda$. In this case, we can Taylor expand the logarithm, considering that $|\xi_\bk|, |\ve| \ll \min\{v_F \Lambda, v_B \Lambda\}$:
\be  
\ln \frac{\xi_\bk-v_FQ - i(\ve + \Omega)}{\xi_\bk+v_FQ - i(\ve + \Omega)} \approx \ln\frac{-v_FQ-i\Omega}{v_FQ-i\Omega} - \frac{2v_FQ(\xi_\bk - i\ve)}{(v_F^2Q^2 + \Omega^2)^2} +\mathcal{O}(\ve^2, \xi_\bk^2, \ve\xi_\bk).
\ee 
Plugging this back into Eq.~\eqref{SMEq:Sigma3Dgen} and calculating the integral in polar coordinates, we find 
\be 
\Sigma_1(i\ve,\xi_\bk) \approx - \frac{\lambda^2 \Lambda^2}{2^8 \pi^4 v_B^2 v_F(v_B + v_F)}(i\ve - \xi_\bk).
\ee 
The second contribution originates in the case of nonzero $\ve$ and comes from the region $\Omega \sim \ve \ll v_F Q \sim v_F \Lambda$. In this case, we can set $\xi_\bk \approx 0$ and write 
\be  
\ln \frac{-v_FQ - i(\ve + \Omega)}{v_FQ - i(\ve + \Omega)} \approx -\pi i \,\text{sign}(\ve + \Omega), 
\ee 
leading to the contribution 
\be  
\Sigma_2(i\ve,\xi_\bk) \approx \frac{\lambda^2 \Lambda^2}{2^8 \pi^5 v_B^3 v_F}i\ve.
\ee 
One can show that the analogous contribution {\it does not} originate in the case of nonzero $\xi_\bk$.

Collecting together, we obtain 
\be  
\Sigma(i\ve,\xi_\bk) = \Sigma_1(i\ve,\xi_\bk) + \Sigma_2(i\ve,\xi_\bk) \approx \frac{\lambda^2 \Lambda^2}{2^8 \pi^4 v_B^3 v_F(v_B + v_F)}(i\ve v_F + \xi_\bk v_B). \label{SMEq:fermionSE3D}
\ee 

We see that the self-energy correction is regular in $\ve$, $\xi_\bk$ and small at weak coupling. The exact numerical coefficient in front of it depends on the details of the ultraviolet regularization. Furthermore, since the relevant frequencies and momenta are $Q \sim |\Omega|/v_B \sim \Lambda$, the logarithmic accuracy in Eq.~\eqref{SMEq:bosonicbubble3D} is not sufficient. However, it does not change the conclusion that the bare fermion self-energy does not destroy the Fermi liquid, as the integral for $\Sigma(i\ve,\xi_\bk)$ does not contain any infrared divergencies. 

Finally, this simple analysis does not take into account the instability of the boson to a new finite-momentum state discussed above. Nevertheless, this result will be of value below, where we consider the transverse boson. 

\subsection{(c) Vertex corrections}

The leading contribution to the bare vertex correction is given by the following diagram: 
\be  
\Gamma(Q,p,k) = \includegraphics[width=2.9cm,valign=c]{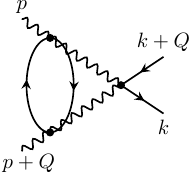} = -\frac{\lambda^3 N}2 \int \frac{d^4l}{(2\pi)^4}\frac{d^4q}{(2\pi)^4} G_0(l)G_0(l+q-p)D_0(q)D_0(q+Q). \label{SMEq:Gamma}
\ee 
Here we used four-vector notations (including frequency) for $l$, $q$, $p$, $k$ and $Q$ for brevity. Once again, two powers of factor $1/2$ from the interaction vertex are canceled by the symmetry factor of the diagram. 

The integration over ``fast'' fermionic momentum $l$ is performed in Eq.~\eqref{SMEq:G0G0bubble3D}. We find for $|Q|, |p|, |k-k_F|\ll \Lambda$
\be  
\Gamma(Q,p,k) \approx \frac{\lambda^3 N N_0}2 \int_{\sim Q}^{\sim \Lambda} \frac{d\omega d^3q}{(2\pi)^4}\frac1{(\omega^2 + v_B^2 q^2)^2}\left\{ 1 - \frac{\omega}{v_Fq} \arctan\left( \frac{v_F q}{\omega} \right)  \right\} \approx \frac{\lambda^3 N N_0 v_F}{16 \pi^2 v_B^3(v_F+v_B)} \ln \frac{\Lambda}{Q}, \label{SMEq:Gamma(Q)}
\ee 
where we used the shorthand notation $Q = \sqrt{|\bQ|^2 + (\Omega/v_B)^2}$. The relevant momenta contributing to this integral are $q \in (Q,\Lambda)$, so we neglected the external momentum $p$ compared to $q$ in the fermionic ``bubble''. We then see that for $p\lesssim Q$ the vertex correction depends on the momentum $|\bQ|$ and frequency $\Omega$ transferred between the two boson (or fermion) lines only, i.e., $\Gamma(Q,p,k) = \Gamma(Q)$. 

Again, this bare vertex correction does not account for the finite-momentum boson instability, so it is not self-consistent. However, this calculation will be valuable when considering the transverse boson in the following section. 

Finally, we estimate another lowest-order vertex correction that contains bosonic ``bubble'' instead of fermionic one. We find that 
\be 
\tilde \Gamma = \includegraphics[width=2.9cm,valign=c]{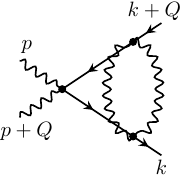} \sim \lambda^3 M \int \frac{d^4l}{(2\pi)^4}\frac{d^4q}{(2\pi)^4} G_0(l)G_0(l+Q)D_0(q)D_0(q+k-l)  \lesssim \frac{\lambda^3 M N_0 \Lambda}{v_B^3 k_F} \ll \Gamma(Q).
\ee 
Indeed, not only does this diagram have a smallness $\Lambda/k_F\ll1$ compared to $\Gamma(Q)$, but it also does not contain a large logarithmic factor $\ln(\Lambda/Q)$, so it can be safely neglected.

\section{II. Calculation in $\bf d=3$ spatial dimensions for a transverse boson}

Now we switch to analyzing the transverse boson as specified in Eq.~\eqref{Eq:phi} of the main text. This modification implies that the number of bosonic field components is equal to the number of spatial dimensions $M=d$, and only the components transverse to their momentum are critical ($d-1$ independent modes). The longitudinal mode remains massive and does not affect the critical properties of the system. The projection onto the transverse sector allows one to stabilize the uniform critical point at a sufficiently large ratio $v_F/v_B$, although we demonstrate that in this case the system flows to the strong-coupling regime at low energies.

\subsection{(a) Perturbative diagrammatic analysis}
The technical implication of having transverse critical modes only is that now the boson propagator, its self-energy, and vertex corrections are matrices in the component space which must be projected onto the transverse sector using the projector $P^T_{ij}(\bQ) = \delta_{ij} - \hat Q_i \hat Q_j$. Here we define the unit vector $\hat \bQ = \bQ/|\bQ|$. The bare boson propagator is then given by the matrix 
\be  
D_{0,ij}^{T}(i\Omega,\bQ) = D_0(i\Omega,\bQ) P^T_{ij}(\bQ) =\frac{\delta_{ij} - \hat Q_i  \hat Q_j}{\Omega^2 + v_B^2 Q^2}. 
\ee 
With this important modification, the boson self-energy now takes the form
\begin{align}  
\Pi_{ij}(i\Omega,\bQ) = \includegraphics[width=2.5cm,valign=c]{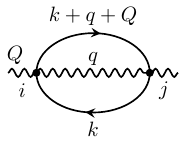} &= \lambda^2 N \int \frac{d\omega d^3\bq}{(2\pi)^{4}}  \frac{d\ve d^3\bk}{(2\pi)^{4}} G_0(i\ve,\bk) G_0(i\ve+i\omega + i\Omega, \bk + \bq + \bQ) D^T_{0,ij}(i\omega, \bq) \nonumber\\&=\Pi_{ij}(0,0) + \Pi_T(i\Omega,\bQ) P_{ij}^T(\bQ) +  \Pi_L(i\Omega,\bQ) P_{ij}^L(\bQ),
\end{align}
where index $L$ stands for ``longitudinal'' and $P_{ij}^L(\bQ) = \hat Q_i \hat Q_j$ is the projector onto the longitudinal mode. Integration over $(\ve,\bk)$ is performed using Eq.~\eqref{SMEq:G0G0bubble3D}.

At zero momentum $\bQ=0$, the angle averaging is given by $\langle \hat q_i \hat q_j  \rangle = \delta_{ij}/3$. This leads to
\be  
\Pi_{ij}(i\Omega,0) = \frac23 \Pi(i\Omega,0) \delta_{ij},
\ee 
which implies that
\be  
\Pi_{ij}(0,0) = \frac23 \Pi(0,0) \delta_{ij}, \qquad \Pi_{T}(i\Omega,0) = \Pi_{L}(i\Omega,0) = \frac23 \delta\Pi(i\Omega,0). \label{SMEq:PiOmegaTL3D}
\ee 
The expressions for $\Pi(i\Omega,0)$, $\Pi(0,0)$, and $\delta\Pi(i\Omega,0)$  are given by Eqs.~\eqref{SMEq:PideltaPi3D}-\eqref{SMEq:Pibare3D}. 

The expression for $\Pi_{ij}(0,\bQ)$ looks more complicated. Projecting it onto the transverse and longitudinal sectors and performing integration, we find for $Q \ll \Lambda$
\begin{align}
&\Pi_{T,L}(0,\bQ) = \frac{\lambda^2 N N_0}{12\pi^2 v_B^3}(v_B Q)^2\left[g_Q^{T,L}\left( \frac{v_F}{v_B} \right)\ln\frac{\Lambda}Q + \tilde g_Q^{T,L}\left( \frac{v_F}{v_B} \right) \right], \label{SMEq:PiQTL3D}\\ &g_Q^{T}(x) = \frac15\left( 
 \frac{\ln(1+x)}x - \frac{1+3x}{(1+x)^2} \right),\qquad g_Q^{L}(x) =\frac15\left( \frac{2+x}{(1+x)^2} - \frac{2\ln(1+x)}x\right).\nonumber
\end{align}
Explicit form of the sublogarithmic terms $\tilde g_Q^{T,L}(x)$ depends on the regularization scheme and is not  important. 

Since we assume in this section that the longitudinal mode remains massive and therefore is away from criticality,  we focus our attention on $\Pi_T$ solely. We see that $g_Q^T(x)$ changes sign at $x^* \approx 13.3$. This indicates that for the ratios $v_F/v_B$ below $x^*$ the system is again unstable to the nonuniform modulated state with the ordering wave vector 
\be  
Q_0^T \sim \Lambda \exp\left( - \frac{12 \pi^2 v_B^3}{\lambda^2 N N_0|g_Q^T(v_F/v_B)|}\right). 
\ee 

For larger values of $v_F/v_B$, on the other hand, the uniform state remains stable, at least within the lowest-order perturbative analysis. To study this state further, we calculate the leading-order fermion self-energy and vertex correction. We find that the fermion self-energy retains the Fermi-liquid form and is given by an expression similar to Eq.~\eqref{SMEq:fermionSE3D}, only with different numerical prefactors. The leading vertex correction also has form similar to Eqs.~\eqref{SMEq:Gamma}-\eqref{SMEq:Gamma(Q)} and is equal to 
\be  
\Gamma_{ij}^T(Q,p,k) =  -\frac{\lambda^3 N}2 \int \frac{d^4l}{(2\pi)^4}\frac{d^4q}{(2\pi)^4} G_0(l)G_0(l+q-p)D^T_{0,il}(q)D_{0,lj}^T(q+Q) \approx  \frac{\lambda^3 N N_0 v_F \delta_{ij}}{24 \pi^2 v_B^3(v_F+v_B)} \ln \frac{\Lambda}{Q} \equiv \Gamma^T(Q) \delta_{ij}, \label{SMEq:GammaT3D}
\ee
where we used $P^T_{lj}(\bq + \bQ)\approx P^T_{lj}(\bq)$ for $Q \ll q \in (Q,\Lambda)$.

If we write bare boson propagator in a more general form $D^{T}(i\Omega,\bQ) =(Z^2 \Omega^2 + v_B^2 Q^2)^{-1}$, then Eqs.~\eqref{SMEq:Pibare3D}, \eqref{SMEq:PiOmegaTL3D}, and~\eqref{SMEq:PiQTL3D} can be viewed as perturbative scale-dependent corrections to $Z$ and $v_B$, while Eq.~\eqref{SMEq:GammaT3D} is the correction to the coupling constant $\lambda$: 
\begin{align}  
&\delta \left(Z^2\right) \approx \frac{\lambda^2 N N_0 Z^2 v_F}{12 \pi^2 v_B^2(Zv_F + v_B)^2}\ln \frac{v_B \Lambda}{Z|\Omega|}, \qquad \delta \left(v_B^2\right) \approx \frac{\lambda^2 N N_0}{12 \pi^2 Z v_B }g_Q^T\left(\frac{Z v_F}{v_B}\right) \ln \frac{\Lambda}Q, \nonumber \\ &\delta\left( \frac{\lambda}2 \right) \approx \Gamma^T(Q) =   \frac{\lambda^3 N N_0 v_F}{24 \pi^2 v_B^3(Z v_F + v_B)}\ln \frac{\Lambda}{\sqrt{Q^2 + Z^2\Omega^2/v_B^2}}. \label{SMEq:RGcorrections}
\end{align}
Here we generalized these expressions for an arbitrary $Z$. Written in this form, these equations allow us to perform calculations beyond the lowest-order corrections by deriving and solving the renormalization group (RG) flow equations in the next section. This approach effectively sums up ``parquet'' diagrams, i.e., those that can be drawn without intersections. It accounts for large logarithmic factors in a controllable way in the weak-coupling regime.


\subsection{(b) Renormalization group equations}

Logarithmic terms in Eqs.~\eqref{SMEq:RGcorrections} that diverge at low energies make this system suitable for study within the RG flow approach. To derive corresponding RG equations, we allow the parameters that have large logarithmic corrections ($Z$, $v_B$, and $\lambda$) to depend on the current energy scale. Fermionic parameters do not acquire such corrections since fermionic self-energy is regular. By performing integration within the small momentum shell at each step of the RG flow, we obtain from Eqs.~\eqref{SMEq:RGcorrections}
\begin{align}
&\frac1{Z^2}\frac{dZ^2}{dl} = \alpha g^T_\Omega\left(\frac{Z v_F}{v_B}\right), \qquad &&g^T_\Omega(x) = \frac{x}{(x+1)^2}, \nonumber \\ &\frac1{v_B^2}\frac{dv_B^2}{dl} = \alpha g^T_Q\left(\frac{Z v_F}{v_B}\right), \qquad &&g^T_Q(x) = \frac15 \left(\frac{\ln(1+x)}x - \frac{1+3x}{(1+x)^2}  \right), \label{SMEq:RGraw} \\ &\frac1{\lambda}\frac{d\lambda}{dl} = \alpha g^T_\lambda\left(\frac{Z v_F}{v_B}\right), \qquad &&g^T_\lambda(x) = \frac{x}{x+1},  \nonumber
\end{align}
where $l = \ln (\Lambda/q)$ is the ``RG time'' and $q$ is the current momentum/frequency scale. Here, we defined new variables
\be  
\alpha = \frac{\lambda^2 N N_0}{12\pi^2 Z v_B^3}, \qquad x = \frac{Z v_F}{v_B},
\ee 
which allows us to rewrite RG equations~\eqref{SMEq:RGraw} in the closed form
\begin{align} 
&\frac1{\alpha} \frac{d\alpha}{dl}  = \alpha \hat g(x), \qquad \hat g(x) = 2 g^T_\lambda(x) - \frac{g^T_\Omega(x)}2 - \frac{3g^T_Q(x)}2 = \frac1{10}\left(\frac{3 + 24x + 20x^2}{(1+x)^2} - \frac{3\ln(1+x)}x\right), \nonumber \\ &\frac1{x}\frac{dx}{dl} = \frac{\alpha}2 \left[g^T_\Omega(x) - g^T_Q(x)\right] = \frac{\alpha}{10}\left( \frac{1+8x}{(1+x)^2} - \frac{\ln(1+x)}x\right).
\end{align}
These equations are valid in the weak-coupling limit $\alpha \ll 1$ and can be easily integrated numerically. However, we notice that in this limit $x(l)$ almost does not change under RG flow because the right-hand side of the corresponding equation is numerically small for all $x$: 
\be 
x(l) \approx x(0) = \text{const}.
\ee 
Under this assumption, we can easily solve the equations for $\alpha$, $Z$, $v_B$, and $\lambda$: 
\begin{align}
&\alpha(l) \approx \frac{\alpha(0)}{1 - \alpha(0) \hat g(x) l}, \qquad &&\lambda(l) \approx \frac{\lambda(0)}{[1-\alpha(0) \hat g(x) l]^{g^T_\lambda/\hat g}}, \nonumber\\ &Z^2(l) \approx \frac{Z^2(0)}{[1-\alpha(0) \hat g(x) l]^{g^T_\Omega/\hat g}}, \qquad &&v_B^2(l) \approx \frac{v_B^2(0)}{[1-\alpha(0) \hat g(x) l]^{g^T_Q/\hat g}}. \label{SMEq:RGsol}
\end{align}
The argument of all $g$-functions is $x = Z v_F/v_B$ which we assume to be a constant, and the natural choice is $Z(0) = 1$. 

This result shows us that the system flows to the strong-coupling regime at low energies. It is reached at the scale given by $\alpha(0) \hat g(x) l \sim 1$, where Eqs.~\eqref{SMEq:RGsol} and the assumption $x(l)\approx \text{const}$ break down. We find numerically that the region where they break down is very narrow. 
At $v_F/v_B > x^*\approx 13.3$, where the uniform quantum critical point (QCP) remains stable, we have $\hat g(x) \approx 2 g^T_\lambda(x) \approx 2$, since $x\gg1$ in this case. However, Eqs.~\eqref{SMEq:RGsol} are correct even at $v_F/v_B < x^*$ such that $g^T_Q(x) < 0$, demonstrating that the bosonic velocity $v_B(l)$ vanishes at the same scale $l \sim 1/\alpha \hat g$ and again indicating the instability toward the finite-momentum state.

\section{III. Calculation in $\bf d=2$ spatial dimensions}

The technical approach to analyzing two-dimensional systems is similar to the case of $d=3$, so we skip some of the calculational details. As expected, we find that the effect of quantum fluctuations in $d=2$ is much stronger and more pronounced compared to $d=3$.

\subsection{(a) Bare boson self-energy}

The bare fermionic ``bubble'' in the limit $q, |\omega|/v_F \ll k_F$ is now given by 
\be  
\int \frac{d\ve d^2\bk}{(2\pi)^{3}} G_0(i\ve,\bk) G_0(i\ve+i\omega, \bk + \bq ) = \includegraphics[width=2.3cm,valign=c]{Bare_Fermion_Bubble} = - N_0\left\{1 - \frac{|\omega|}{\sqrt{\omega^2 + v_F^2 q^2}}\right\}, \label{SMEq:G0G0bubble2D}
\ee 
leading to the lowest-order nontrivial bare boson self-energy 
\be
\Pi(i\Omega,\bQ) = \includegraphics[width=2.5cm,valign=c]{2-Loop_Bare_Boson_Self-Energy} = - \lambda^2 N N_0 \int \frac{d\omega d^2\bk}{(2\pi)^{3}}  \left(1 - \frac{|\omega + \Omega|}{\sqrt{(\omega+\Omega)^2 + v_F^2 (\bk + \bQ)^2}}\right)\frac1{\omega^2 + v_B^2 k^2}.
\ee
We again separate out the constant and frequency/momentum-dependent terms, $\Pi(i\Omega,\bQ) = \Pi(0,0) + \delta \Pi(i\Omega,\bQ)$. We find that 
\be  
\Pi(0, 0) = -\lambda^2 N N_0 \int \frac{d\omega d^2\bk}{(2\pi)^{3}}  \left(1-\frac{|\omega|}{\sqrt{\omega^2 + v_F^2 k^2}}\right)\frac1{\omega^2 + v_B^2 k^2} = -\frac{\lambda^2 N_0 N \Lambda}{2\pi v_B}  \left\{\frac12 - \frac{\text{ArcCosh}(v_F/v_B)}{\pi \sqrt{(v_F/v_B)^2-1}}  \right\}, \label{SMEq:Pi00}
\ee 
where the analytical form of this function should be understood as
\be  
\frac12 - \frac{\text{ArcCosh}(x)}{\pi \sqrt{x^2-1}} = \left\{ \begin{array}{cc} \displaystyle \frac12 - \frac{\text{ArcCos}(x)}{\pi \sqrt{1-x^2}} , & x<1 \\[1.2em] \displaystyle \frac12 - \frac{\text{ArcCosh}(x)}{\pi \sqrt{x^2-1}} , & x>1 \end{array}     \right. \approx \left\{ \begin{array}{cc} \displaystyle \frac{x}{\pi} , & x\ll1 \\[1.2em] \displaystyle \frac12 - \frac{\ln(2x)}{\pi x} , & x\gg1. \end{array}     \right. 
\ee 
This term does not have any infrared divergencies and only shifts the position of the critical point. 

The frequency and momentum dependence of $\Pi$ is determined by the expression
\be  
\delta\Pi(i\Omega, \bQ) = - \lambda^2 N N_0 \int \frac{d\omega d^2\bk}{(2\pi)^{3}}  \left(\frac{|\omega|}{\sqrt{\omega^2 + v_F^2 k^2}} - \frac{|\omega + \Omega|}{\sqrt{(\omega+\Omega)^2 + v_F^2 (\bk + \bQ)^2}}\right)\frac1{\omega^2 + v_B^2 k^2}. 
\ee
At finite frequency, we obtain
\begin{align}  
&\delta \Pi(i\Omega,0) =  \frac{\lambda^2 N N_0}{v_B^2}f_\Omega\left( \frac{v_F}{v_B}\right) |\Omega|, \label{SMEq:fW2DnonT}\\ &f_\Omega(x) = \frac{1}{2\pi^2} \int_0^{\infty} \frac{t^2 \, dt}{t^2 + x^2}\left(\frac1{1+t^2}\right)^{3/2} \left\{ t \ln \left(t+\sqrt{t^2+1} \right) - \sqrt{1+t^2}  \right\} \approx   \frac1{4\pi}\left\{ \begin{array}{cc} x, & x\ll1 \\[.8em] \displaystyle \frac{\ln(2x)-1}{x}, & x\gg1. \end{array} \right.  \nonumber
\end{align}
At finite momentum, we find
\begin{align}  
&\delta \Pi(0,\bQ) = \frac{\lambda^2 N N_0}{v_B}f_Q\left( \frac{v_F}{v_B}\right) Q, \label{SMEq:fQ2DnonT}\\ &f_Q(x) = - \frac{x}{2\pi^2} \cdot \left\{ \begin{array}{cc} \displaystyle\frac1{x^2-1} + \frac{\text{ArcTanh}\left(\sqrt{1-x^2}\right)}{(1-x^2)^{3/2}} , & x<1 \\[1.2em] \displaystyle\frac1{x^2-1} - \frac{\text{ArcTan}\left(\sqrt{x^2-1}\right)}{(x^2-1)^{3/2}} , & x>1 \end{array}     \right. \approx - \frac1{2\pi^2}\left\{ \begin{array}{cc}\displaystyle x\left(\ln\frac{2}{x} - 1  \right) , & x\ll1 \\[1.2em] \displaystyle\frac1x , & x\gg1. \end{array}     \right.  \nonumber
\end{align}
After summing up, these expressions reproduce Eq.~\eqref{Eq:Pi2D} of the main text. Functions $f_\Omega(x)$ and $f_Q(x)$ are shown in Fig.~\ref{SMFig:f}(a). 

We stress that $f_Q(x) < 0$ for all $x$, indicating the instability toward the nonuniform finite-momentum state with the wave vector
\be  
Q_0 = \frac{\lambda^2 N N_0}{2v_B^3} \left| f_Q\left( \frac{v_F}{v_B}\right)\right|.
\ee 
\begin{figure}
 \begin{center}
   \includegraphics[width=1.\linewidth]{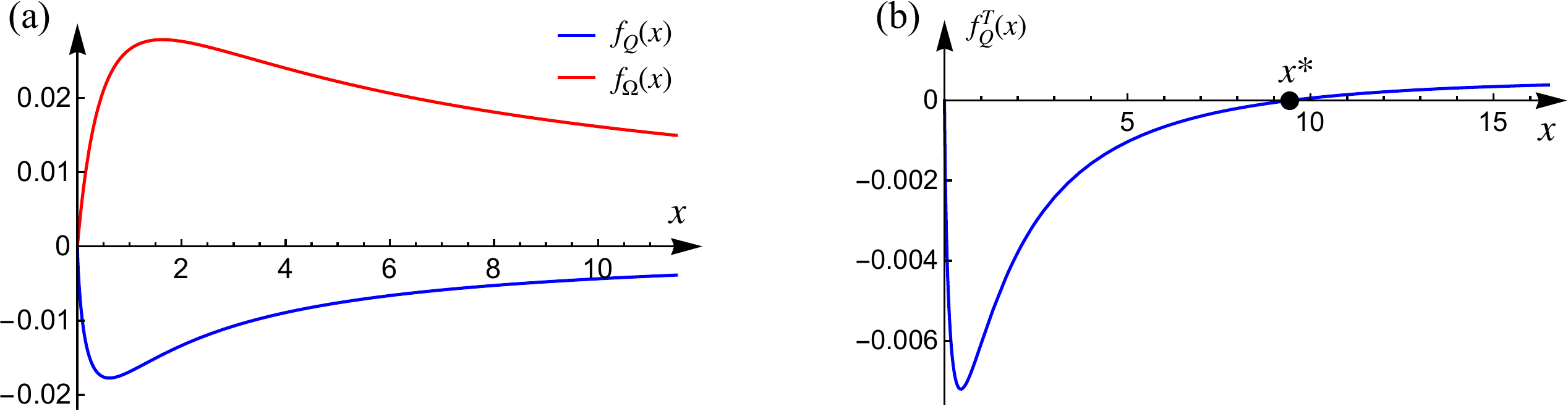}
 \end{center}
\caption{(a) Functions $f_\Omega(x)$ and $f_Q(x)$ for an isotropic boson, defined in Eqs.~\eqref{SMEq:fW2DnonT} and~\eqref{SMEq:fQ2DnonT}. Negative sign of $f_Q(x)$ implies the instability toward the finite-momentum state at any ratio $v_F/v_B$. (b) Function $f^T_Q(x)$ for a transverse critical boson, given by Eq.~\eqref{SMEq:PiQTL2D}. It changes sign at $x^*\approx 9.43$.}
 \label{SMFig:f}
\end{figure}

\subsection{(b) Bare fermion self-energy}

To make our study comprehensive, we present the calculation for the bare fermion self-energy, even though such a calculation is not self-consistent because of the bosonic finite-momentum instability. 

The bare bosonic ``bubble'' equals
\be  
\int \frac{d\omega d^2\bq}{(2\pi)^{3}} D_0(i\omega, \bq)  D_0(i\omega + i\Omega, \bq + \bQ) = \includegraphics[width=2.3cm,valign=c]{Bare_Boson_Bubble} = \frac1{8v_B^2 \sqrt{\Omega^2 + v_B^2 Q^2}},
\ee 
leading to fermion self-energy (with logarithmic accuracy)
\begin{align}
\Sigma(i\ve,\xi_\bk) = \includegraphics[width=2.5cm,valign=c]{2-Loop_Bare_Fermion_Self-Energy} &= i\frac{\lambda^2 M}{64\pi^2 v_B^2} \int_0^{\infty}Q \, dQ \int_{-\infty}^{\infty} d\Omega \frac{\text{sign}(\ve+\Omega)}{\sqrt{v_F^2  Q^2 + (\ve+ \Omega+ i \xi_\bk)^2}} \cdot \frac1{\sqrt{\Omega^2 + v_B^2 Q^2}} \approx \nonumber \\ & \approx \frac{\lambda^2 M}{32 \pi^2 v_B^3 v_F (v_B + v_F)} \left(i \ve v_F + \xi_\bk v_B \right) \ln \left( \frac{\Lambda v_B}{\max\{|\ve|, |\xi_\bk|  \}}  \right). \label{SMEq:electronSE2Dbare}
\end{align}
It has the form of a marginal Fermi liquid. However, since the boson propagator in this calculation is the bare one and does not account for the finite-momentum instability, we cannot draw any reliable conclusions about the marginal-Fermi-liquid behavior based on this result. In fact, as we show below, the {\it self-consistent uniform (zero-momentum)} solution for the transverse boson preserves Fermi liquid.

\subsection{(c) Bare vertex correction}
The leading bare vertex correction is given by 
\be  
\Gamma(Q,p,k) = \includegraphics[width=2.9cm,valign=c]{Bare_Fermion_Vertex_Correction} = \frac{\lambda^3 N N_0}2 \int \frac{d\omega d^2q}{(2\pi)^3}\frac{1 - \frac{|\omega-\omega_p|}{\sqrt{(\omega-\omega_p)^2 + v_F^2(\bq - \bp)^2}}}{\left[ \omega^2 + v_B^2 q^2 \right]\left[(\omega+\Omega)^2 + v_B^2 (\bq+\bQ)^2\right]}.
\ee 
This expression can be most easily evaluated in the limit $v_F \gg v_B$. Using the Feynman parameterization, we find
\be
\Gamma(Q,p,k) \approx \frac{\lambda^3 N N_0}{16 v_B^2 \sqrt{\Omega^2 + v_B^2 Q^2}}. \label{SMEq:vertexbare12D}
\ee
Strong infrared divergence of the vertex correction at small scales is an artifact of using bare boson propagators, again. As we demonstrate in the next section, when studying uniform QCP, a calculation with the self-consistent solution replaces this power-law divergence with the logarithm. 

Another lowest-order bare vertex correction can be roughly estimated as
\be 
\tilde \Gamma = \includegraphics[width=2.9cm,valign=c]{Bare_Boson_Vertex_Correction} \lesssim \frac{\lambda^3 M N_0}{v_B^3 k_F} \ln\frac{\Lambda}{\max\{Q,k\}} \ll \Gamma(Q,p,k). \label{SMEq:vertexbare22D}
\ee

\section{IV. Calculation in $\bf d=2$ spatial dimensions for a transverse boson}

The key effects arising from the transverse nature of the critical boson field, Eq.~\eqref{Eq:phi}, are qualitatively similar to those in the $d=3$ case. The boson propagator, its self-energy, and the vertex corrections must all be projected onto the transverse sector using the projector $P_{ij}^T(\bQ) = \delta_{ij} - \hat Q_i \hat Q_j$. As in $d=3$, the uniform critical point can be stabilized if the ratio $v_F/v_B$ is sufficiently large, in which case the system enters a strong-coupling regime at low energies. However, unlike the $d=3$ case, where corrections are logarithmic, here they follow a power-law scaling. Consequently, we find that the boson field acquires a nonzero anomalous dimension, $\eta = 1/2$.

\subsection{(a) Bare self-energies and vertex corrections}

Bare {\it fermion self-energy} and {\it vertex corrections} are given by expressions similar to Eqs.~\eqref{SMEq:electronSE2Dbare}, \eqref{SMEq:vertexbare12D}, and~\eqref{SMEq:vertexbare22D}, only with different numerical coefficients. 

The {\it boson self-energy} looks more sophisticated now:
\begin{align}  
\Pi_{ij}(i\Omega,\bQ) = \includegraphics[width=2.5cm,valign=c]{2-Loop_Bare_Transverse_Boson_Self-Energy} &= \lambda^2 N \int \frac{d\omega d^2\bq}{(2\pi)^{3}}  \frac{d\ve d^2\bk}{(2\pi)^{3}} G_0(i\ve,\bk) G_0(i\ve+i\omega + i\Omega, \bk + \bq + \bQ) D^T_{0,ij}(i\omega, \bq) \nonumber\\&=\Pi_{ij}(0,0) + \Pi_T(i\Omega,\bQ) P_{ij}^T(\bQ) +  \Pi_L(i\Omega,\bQ) P_{ij}^L(\bQ). \label{SMEq:Pitrans2D}
\end{align}
Integration over fermionic frequency/momentum $(\ve, \bk)$ is performed using Eq.~\eqref{SMEq:G0G0bubble2D}. Angle averaging at $\bQ=0$ is given now by $\langle \hat q_i \hat q_j \rangle = \delta_{ij}/2$, leading to
\be  
\Pi_{ij}(0,0) = \frac12 \Pi(0,0) \delta_{ij}, \qquad \Pi_{T}(i\Omega,0) = \Pi_{L}(i\Omega,0) = \frac12 \delta\Pi(i\Omega,0), \label{SMEq:PiOmegaTL2D}
\ee 
where $\Pi(0,0)$ and $\delta\Pi(i\Omega,0)$ are given by Eqs.~\eqref{SMEq:Pi00}-\eqref{SMEq:fW2DnonT}.

At finite momentum, we find
\begin{align}
&\Pi_{T,L}(0,\bQ) = \frac{\lambda^2 N N_0}{v_B^2}f_Q^{T,L}\left( \frac{v_F}{v_B} \right)v_B Q, \label{SMEq:PiQTL2D}\\ &f_{Q}^T(x)  = \frac1{12 \pi^2 x} \int_0^1 \frac{dt}t \left\{ 1 - \sqrt{1-t}\frac{1-t+3 t x^2}{(1-t + t x^2)^{3/2}} \right\} \approx \frac{1}{24 \pi^2}\left\{ \begin{array}{cc} \displaystyle - x\left( 6 \ln \frac2x - 7\right) + O(x^3\ln x), & x\ll1, \\[1.em] \displaystyle\frac4x \left( \ln\frac{x}2 - 2 \right) + O \left(  \frac1{x^2}\right), & x\gg1, \end{array}     \right. \nonumber \\[.8em] &f_{Q}^L(x)  = \frac1{12 \pi^2 x} \int_0^1 \frac{dt}t \left\{ \frac{(1-t)^{3/2}}{(t x^2 + 1 - t)^{3/2}} - 1 \right\} \approx -\frac{1}{24 \pi^2}\left\{ \begin{array}{cc} \displaystyle x\left( 6 \ln \frac2x - 5\right) + O(x^3\ln x), & x\ll1, \\[1.em] \displaystyle\frac4x \left( \ln\frac{x}2 +1 \right) + O \left(  \frac1{x^3}\right), & x\gg1. \end{array}     \right. \nonumber
\end{align}
Function $f_Q^T(x)$ is shown in Fig.~\ref{SMFig:f}(b). It changes sign at $x^* \approx 9.43$, indicating the finite-momentum instability if ratio $v_F/v_B$ is below this value, with the ordering wave vector 
\be  
Q_0^T = \frac{\lambda^2 N N_0}{2v_B^3} \left|f^T_Q\left( \frac{v_F}{v_B}\right)\right|.
\ee 
At ratios above $x^*$, the uniform zero-momentum QCP remains stable within the perturbative analysis. However, since the self-energy correction is linear in $Q$ and $\Omega$, it dominates over the bare boson propagator below the scale $Q_0\sim Q^T_{0}\sim\lambda^2 N N_0/v_B^3$. This means that the perturbative calculation, Eq.~\eqref{SMEq:Pitrans2D}, is not self-consistent below this scale, as it uses the bare boson propagator $D_0$. In the following section, we discuss the self-consistent low-energy solution for the transverse boson at the uniform QCP.

\subsection{(b) Self-consistent solution neglecting vertex corrections}

The scaling form of the self-consistent solution at low energies can be derived from a simple power counting. This analysis dictates that the solution must scale as $D_T^{-1} = \Pi_T \propto \Omega^{3/2}, Q^{3/2}$ at low frequencies and momenta, provided that such a solution exists. We search for it using the following ansatz: 
\be  
D_T^{-1}(i\Omega,\bQ) \approx \Pi_T(i\Omega, \bQ) \approx \Omega_0^{1/2}\left(\Omega^2 + c^2 Q^2  \right)^{3/4}, \label{SMEq:DTSC}
\ee 
where parameters $\Omega_0$ and $c$ must be determined self-consistently. This particular functional form of the ansatz is chosen for calculational convenience. 

Neglecting vertex corrections for now, the lowest-order self-consistent boson self-energy takes form
\be 
\Pi_{ij}(i\Omega,\bQ) = \includegraphics[width=2.5cm,valign=c]{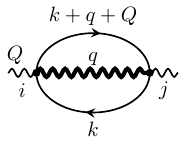} = \lambda^2 N \int \frac{d\omega d^2\bq}{(2\pi)^{3}}  \frac{d\ve d^2\bk}{(2\pi)^{3}} G(i\ve,\bk) G(i\ve+i\omega + i\Omega, \bk + \bq + \bQ) D_T(i\omega, \bq) P^T_{ij}(\bq).
\ee
The bold wavy line represents the dressed boson propagator given by Eq.~\eqref{SMEq:DTSC}. As we demonstrate below, the fermion self-energy calculated self-consistently is regular and retains the Fermi-liquid form. Hence, we can use $G(i\ve,\bk) = G_0(i\ve,\bk)$, assuming that the Fermi-liquid parameters are renormalized. The integral over $\ve$ and $\bk$ can then be performed using Eq.~\eqref{SMEq:G0G0bubble2D}, and we obtain
\begin{align}
\Pi_{ij}(i\Omega, \bQ) - \Pi_{ij}(0,0) &= \lambda^2 N N_0 \int \frac{d\omega d^2 \bk}{(2\pi)^3}\left(\frac{|\omega + \Omega|}{\sqrt{(\omega+\Omega)^2 + v_F^2 (\bk + \bQ)^2}} - \frac{|\omega|}{\sqrt{\omega^2 + v_F^2 k^2}} \right)\frac{\delta_{ij} - \hat k_i \hat k_j}{\Omega_0^{1/2}(\omega^2 + c^2 k^2)^{3/4}} \nonumber \\  &=  \Pi_T(i\Omega, \bQ) P^T_{ij}(\bQ) + \Pi_L(i\Omega, \bQ) P^L_{ij}(\bQ) . \label{SMEq:PiSC}
\end{align}
Using Feynman parameters, we find
\be  
\Pi_T(i\Omega,0) = \Pi_L(i\Omega,0) = \frac{\lambda^2 N N_0}{\Omega_0^{1/2} c^2} |\Omega|^{3/2} \frac{\Gamma(5/4)}{2 \pi^{7/2}\Gamma(3/4)} F_{\Omega}\left( \frac{v_F}c  \right), \qquad \text{where} 
\ee 
\begin{align}  
F_\Omega(x) & = \frac{\pi}{2\sqrt{x}}\int_0^1 \frac{dt}{t^{1/2} (1-t)^{1/4}}  \int_{-\infty}^{\infty} dy \left\{ \frac{|y+1|}{\left[ t (y+1)^2 + (1-t)y^2 x^2  \right]^{1/4}} - \frac{|y|}{\left[ t y^2 + (1-t)y^2 x^2  \right]^{1/4}}  \right\}  \nonumber \\ &\approx \frac{2 \pi^2}3\left\{ \begin{array}{cc} x + O(x^{3/2}), & x\ll1, \\[.5em] \displaystyle \frac2x + O \left(  \frac1{x^{3/2}}\right), & x\gg1, \end{array}     \right.   \label{SMEq:FOmega}
\end{align}
and $\Gamma(x)$ is the gamma-function. At finite momentum, we project $\Pi_{ij}$ onto the transverse sector and obtain 
\be  
\Pi_T(0,\bQ) = \frac{\lambda^2 N N_0}{\Omega_0^{1/2}c^{1/2}}Q^{3/2}\frac{\Gamma(5/4)}{2\pi^{7/2}\Gamma(3/4)} F^T_{Q}\left(\frac{v_F}c\right), \qquad \text{with}
\ee 
\begin{align}  
F^T_{Q}(x) & = \int_0^1 \frac{dt}{t^{1/2} (1-t)^{1/4} (t x^2 + 1-t)^{1/4}} \int_0^{\infty}q dq \int_0^{2\pi} d\varphi \left\{\frac1{\left[q^2 + \frac{tx^2 (2q\cos \varphi+1)}{tx^2 + 1-t}   \right]^{1/4}}   - \frac1{q^{1/2}} \right\}\cos^2\varphi  \label{SMEq:FQT} \\ &\approx  \left\{ \begin{array}{cc} - 6.28x , & x\ll1 \\[.2em] \displaystyle -\frac{\sqrt{2}\pi}{\sqrt{x}} \left(\frac{2\pi}3 - \frac{8\sqrt{2}}{21\pi}G\left[ \left\{\left\{\frac34,1  \right\},\left\{ \frac32  \right\}  \right\},\left\{ \left\{\frac14,\frac12,\frac52   \right\},\left\{  \right\}  \right\},1 \right]   \right), & x\gg1 \end{array}     \right.   \approx \left\{ \begin{array}{cc} - 6.28 x , & x\ll1 \\[.8em]  \displaystyle\frac{5.317}{\sqrt{x}}, & x\gg1, \end{array}     \right.   \nonumber
\end{align}
where $G$ is the Meijer G-function. Functions $F_\Omega(x)$ and $F^T_Q(x)$ are shown in Fig.~\ref{SMFig:F}(a), and $F^T_Q(x)$ changes sign at $x^* \approx 1.97$
\begin{figure}
 \begin{center}
   \includegraphics[width=1.\linewidth]{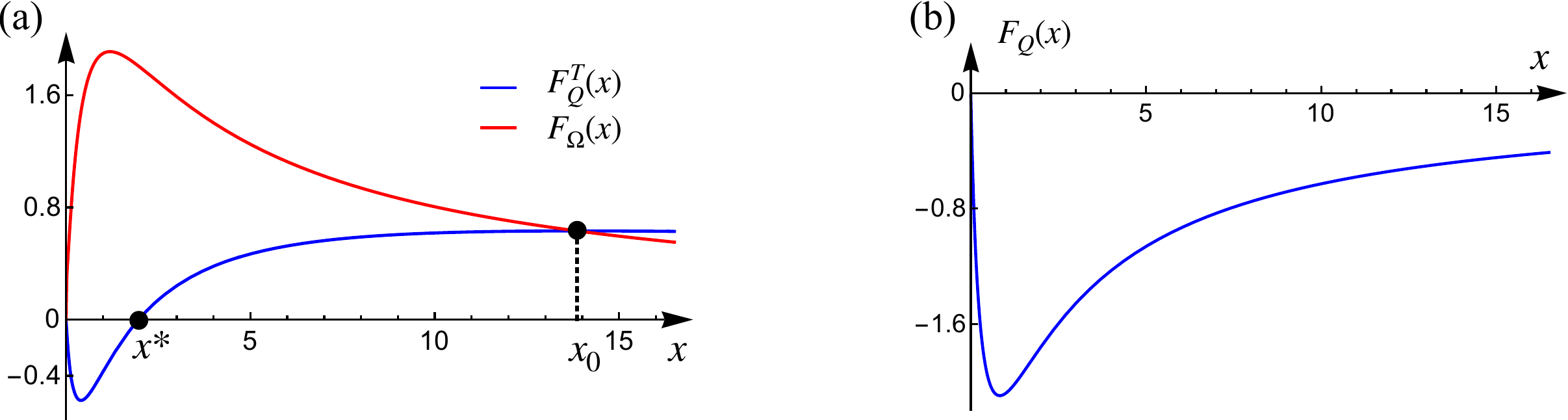}
 \end{center}
\caption{(a) Functions $F_\Omega(x)$ and $F_Q^T(x)$, defined in Eqs.~\eqref{SMEq:FOmega} and~\eqref{SMEq:FQT}, intersect at $x_0 \approx 13.87$ indicating the self-consistent solution for renormalized bosonic velocity $c$.  $F_Q^T(x)$ changes sign at $x^* \approx 1.97$. (b) Function $F_Q(x)$ for an isotropic (not projected onto transverse sector) critical boson given by Eq.~\eqref{SMEq:FQ}. Since $F_Q(x) < 0$ for all $x$, a self-consistent solution of the form~\eqref{SMEq:DTSC}  does not exist in this case.}
 \label{SMFig:F}
\end{figure}

The self-consistency conditions require that $D_T^{-1}(i\Omega,0) = \Pi_T(i\Omega,0)$ and $D_T^{-1}(0,\bQ) = \Pi_T(0,\bQ)$, leading to
\be  
\left\{ \begin{array}{c}\Omega_0^{1/2} = \displaystyle \frac{\lambda^2 N N_0}{\Omega_0^{1/2} c^2}\frac{\Gamma(5/4)}{2 \pi^{7/2}\Gamma(3/4)} F_{\Omega}\left( \frac{v_F}c  \right) \\[1.5em] \displaystyle \Omega_0^{1/2} c^{3/2} =\frac{\lambda^2 N N_0}{\Omega_0^{1/2}c^{1/2}}\frac{\Gamma(5/4)}{2\pi^{7/2}\Gamma(3/4)} F^T_{Q}\left(\frac{v_F}c\right)\end{array}\right. \qquad \Longrightarrow\qquad F_{\Omega}\left( \frac{v_F}c  \right) = F^T_{Q}\left( \frac{v_F}c  \right).
\ee 
Functions $F_\Omega(x)$ and $F_Q^T(x)$ intersect at $x_0 \approx 13.87 > x^*$, indicating that the self-consistent solution exists. Resolving then these equations with respect to $\Omega_0$ and $c$, we find 
\be 
\Omega_0 \approx 0.818 \frac{\lambda^2 N N_0}{v_F^2} , \qquad c = v_F/x_0 \approx 0.072v_F. \label{SMEq:Omega0c}
\ee 
Self-consistent solution~\eqref{SMEq:DTSC} is valid below the scale $\Omega_0$,
\be  
\max\left\{|\Omega|, cQ\right\} \ll \Omega_0,
\ee 
where the self-consistent self-energy~\eqref{SMEq:PiSC} dominates over the bare boson propagator $D_0(i\Omega,\bQ)$.

We also note that our analysis is not entirely rigorous, as we calculated the frequency and momentum dependence of $\Pi_T(i\Omega,\bQ)$ separately. Consequently, the functional form of Eq.~\eqref{SMEq:DTSC} remains our guess, while a more rigorous calculation should be performed numerically. We leave this task for future study. However, as we demonstrate below in Section IV(d) and as is shown in Fig.~\ref{SMFig:F}(b), even within our simplified analysis, there is no solution for the case when the boson is isotropic, i.e., not projected onto the transverse sector. In the latter case, $F_Q(x) < 0$ for all $x$, which is consistent with the bare perturbative analysis. 

The fermion self-energy has a form similar to Eq.~\eqref{SMEq:electronSE2Dbare}. However, the effect of the renormalized boson is to cut off the logarithmic singularity at the scale $\Omega_0$ instead of $\max\left\{ |\ve|, |\xi_\bk|\right\}$. Indeed, the bosonic ``bubble'' can now be estimated as
\be  
\int \frac{d\omega d^2\bq}{(2\pi)^{3}} D_T(i\omega, \bq)  D_T(i\omega + i\Omega, \bq + \bQ) = \includegraphics[width=2.3cm,valign=c]{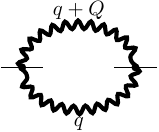} \sim \left\{ \begin{array}{ll} \displaystyle\frac1{c^2\Omega_0}\ln\frac{\Omega_0}{\sqrt{\Omega^2 + c^2 Q^2}}, &\sqrt{\Omega^2 + c^2 Q^2}\ll \Omega_0 \\[1.5em]  \displaystyle \frac1{v_B^2 \max\left\{ |\Omega|, v_B Q \right\}}, & \Omega_0 \ll \max\left\{ |\Omega|, v_B Q \right\} \ll v_B \Lambda. \end{array}     \right. 
\ee 
Then, at low frequencies and momenta close to the Fermi surface,  $\max\{|\ve|,|\xi_\bk|\}< \Omega_0$, the leading contribution to the fermion self-energy comes from momenta $\Omega_0/v_B < q < \Lambda$. The resulting expression is given by
\be 
\Sigma(i\ve,\xi_\bk) = \includegraphics[width=2.5cm,valign=c]{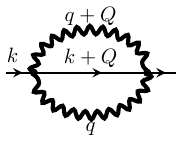} \propto \lambda^2 (i\ve + \xi_\bk) \ln \frac{v_B \Lambda}{\Omega_0}, \qquad  \max\{|\ve|,|\xi_\bk|\}< \Omega_0. \label{SMEq:fermionSESC2D}
\ee
This expression should not be considered literally in the sense that the coefficients in front of $i\ve$ and $\xi_\bk$ might differ. We also dropped all the numerical prefactors in this expression, as well as the factors of $v_F$, $v_B$ and $c$.

Equation~\eqref{SMEq:fermionSESC2D} demonstrates that the fermion self-energy is regular at low energies if the self-consistent transverse boson~\eqref{SMEq:DTSC} is used, thus preserving the Fermi-liquid behavior. This result justifies our earlier assumption that we can use Eq.~\eqref{SMEq:G0D0} for fermion propagator in our self-consistent calculations, as long as the Fermi-liquid parameters are renormalized. The logarithmic vertex corrections we discuss in the next section change numerical prefactors only.


\subsection{(c) Self-consistent solution: the role of vertex corrections}

Now we demonstrate that the vertex corrections drive the uniform QCP into the strong-coupling regime at low energies. The leading-order diagram with the renormalized boson~\eqref{SMEq:DTSC} equals
\be  
\Gamma^T_{ij}(Q,p,k) = \includegraphics[width=2.9cm,valign=c]{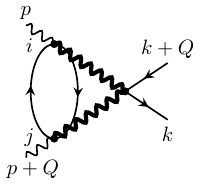} = -\frac{\lambda^3 N}2 \int \frac{d^3l}{(2\pi)^3}\frac{d^3q}{(2\pi)^3} G_0(l)G_0(l+q-p)D_T(q)D_T(q+Q)P^T_{il}(\bq)P^T_{lj}(\bq + \bQ), \label{SMEq:GammaTdiag}
\ee 
where summation over $l$ is implied. According to our discussion above, the dressed fermion propagator has the same form as the bare one but with the renormalized Fermi-liquid parameters.

Assuming that all external frequencies and momenta are small, $\max\{\Omega, c Q, \omega_p, c p, \omega_k, \xi_\bk \} \ll \Omega_0$, we find that the relevant bosonic frequencies and momenta contributing to this integral lie in the range $\sqrt{c^2 Q^2 + \Omega^2} < cq, \,\omega_q < \Omega_0$, leading to 
\be  
\Gamma^T_{ij}(Q,p,k) \approx \frac{\lambda^3 N N_0}{8 \pi^2 \Omega_0 c^2} \frac{v_F}{v_F + c} \ln\frac{\Omega_0}{\sqrt{c^2 Q^2 + \Omega^2}} \delta_{ij} \approx  2.79 \lambda \ln\frac{\Omega_0}{\sqrt{c^2 Q^2 + \Omega^2}} \delta_{ij}, \label{SMEq:GammaTSC2D}
\ee 
where we used the result of Eq.~\eqref{SMEq:Omega0c}, and $\Omega$ is the frequency difference between the external bosonic lines. The contribution from the range $\Omega_0 < cq, \,\omega_q < v_B \Lambda$ is subleading as it does not produce a logarithmic factor. The contribution from the second diagram can be estimated as negligible compared to $\Gamma^T_{ij}$:
\be 
\tilde \Gamma^T_{ij}= \includegraphics[width=2.6cm,valign=c]{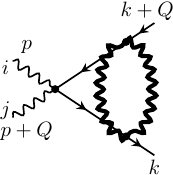} \lesssim \frac{\lambda^3N_0}{v_B^3 k_F} \ln\frac{v_B\Lambda}{\Omega_0} \delta_{ij} \ll \Gamma^T_{ij}(Q,p,k).
\ee 
Equation~\eqref{SMEq:GammaTSC2D} indicates that the system enters the strong-coupling regime below the scale $\Omega_0$, as the vertex correction becomes comparable to or even larger than the bare coupling $\lambda$. Our analytical approach does not allow us to describe this regime in a controllable way. 

To speculate on what the low-energy solution at energies below $\Omega_0$ may look like and capture leading logarithmic corrections, we derive and solve the set of self-consistent equations for the boson propagator and the effective vertex. We confirm that the fermion propagator retains its Fermi-liquid form a posteriori. We define the effective (fully renormalized) boson propagator $D_{\text{eff}}(i\Omega,\bQ)$ and the effective vertex $\Gamma_{\text{eff}}(i\Omega,\bQ)$:
\be  
\includegraphics[width=1.6cm,valign=c]{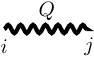} = D_{\text{eff}}(i\Omega,\bQ) P^T_{ij}(\bQ) \qquad \qquad \qquad \includegraphics[width=2.4cm,valign=c]{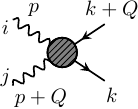} = \Gamma_{\text{eff}}(i\Omega,\bQ) \delta_{ij}.
\ee
We use the insight from Eq.~\eqref{SMEq:GammaTSC2D} and assume that $\Gamma_{\text{eff}}$ depends only on the frequency and momentum difference between the external bosonic lines, $\Omega$ and $\bQ$.

The self-consistent equation for the boson propagator and its self-energy $\Pi_{\text{eff}}$ takes form 
\begin{align}  
&D^{-1}_{\text{eff}}(i\Omega,\bQ)  +\Pi_{\text{eff}}(0,0)  \approx \Pi_{\text{eff}}(i\Omega,\bQ) = \includegraphics[width=2.5cm,valign=c]{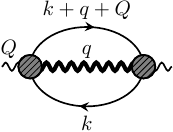} =  \label{SMEq:Deffeq}\\ &= 4 N P^T_{ji}(\bQ)\int \frac{d\omega d^2\bq}{(2\pi)^{3}}  \frac{d\ve d^2\bk}{(2\pi)^{3}} G_0(i\ve,\bk) G_0(i\ve+i\omega + i\Omega, \bk + \bq + \bQ) D_{\text{eff}}(i\omega, \bq) P^T_{ij}(\bq) \Gamma^2_{\text{eff}}\left[i(\omega - \Omega), \bq-\bQ\right] = \nonumber \\ &= -4 N N_0 \int \frac{d\omega d^2\bq}{(2\pi)^{3}}  D_{\text{eff}}(i\omega, \bq) \left( \hat \bq \cdot \hat \bQ\right)^2 \Gamma^2_{\text{eff}}\left[i(\omega - \Omega), \bq-\bQ\right] \left(1 - \frac{|\omega + \Omega|}{\sqrt{(\omega+\Omega)^2 + v_F^2(\bq + \bQ)^2}}\right). \nonumber
\end{align}
We multiplied the equation by $P^T_{ji}(\bQ)$ to project it onto the transverse sector, and then used the equality $P^T_{ij}(\bq) P^T_{ji}(\bQ) = \left(\hat \bq \cdot \hat \bQ \right)^2$. Integration over $(\ve,\bk)$ is performed using Eq.~\eqref{SMEq:G0G0bubble2D}.  We added the constant term $\Pi_{\text{eff}}(0,0)$ to the left-hand side of the equation to account for the shift of the bare critical point.

To derive the equation for the effective vertex $\Gamma_{\text{eff}}(Q)$, we sum up the diagrams without intersections (``parquet'' diagrams) of the form shown in Eq.~\eqref{SMEq:GammaTdiag}, as they contain a large logarithmic factor. Diagrammatically, the summation looks as follows: 
\be  
\includegraphics[width=17cm,valign=c]{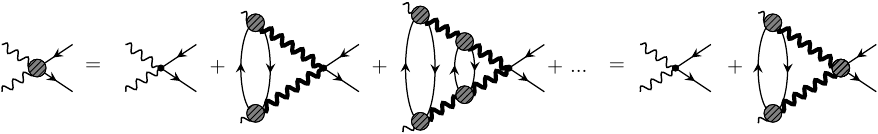} \nonumber
\ee 
The corresponding integral equation for $\Gamma_{\text{eff}}(Q)$ then takes the form:
\begin{align}
\Gamma_{\text{eff}}(Q)\delta_{ij} &\approx \frac{\lambda}2 \delta_{ij} - 4N\int \frac{d^3l}{(2\pi)^3}\frac{d^3q}{(2\pi)^3} G_0(l)G_0(l+q-p)D_{\text{eff}}(q)P_{il}^T(\bq)D_{\text{eff}}(q+Q)P_{lj}^T(\bq+\bQ) \Gamma_{\text{eff}}^2(q-p) \Gamma_{\text{eff}}(Q) \nonumber \\ &\approx \frac{\lambda}2 \delta_{ij} + 2N N_0 \Gamma_{\text{eff}}(Q) \delta_{ij} \int \frac{d\omega d^2 \bq}{(2\pi)^3} D_{\text{eff}}(q)D_{\text{eff}}(q+Q) \Gamma_{\text{eff}}^2(q) \left( 1 - \frac{|\omega|}{\sqrt{\omega^2 + v_F^2 q^2}} \right). \label{SMEq:Gammaeffeq}
\end{align}
Here we put $p=0$ for simplicity, and 4 is the combinatorial factor. We also assumed that $Q \ll q$, which will be justified later. Under this assumption, one finds with logarithmic accuracy that $P_{il}^T(\bq)P_{lj}^T(\bq+\bQ) \approx P_{il}^T(\bq)P_{lj}^T(\bq) = P_{ij}^T(\bq) \to \delta_{ij}/2$.

We look for a low-energy solution in the form 
\be  
D_{\text{eff}}(i\Omega,\bQ) = \frac{\log\frac{\tilde\Omega_0}{\sqrt{\Omega^2 + c^2 Q^2}}}{\tilde\Omega_0^{1/2}\left(  \Omega^2 + c^2 Q^2\right)^{3/4}}, \qquad \Gamma_{\text{eff}}(i\Omega,\bQ) = \frac{\tilde \lambda}{2\log\frac{\tilde\Omega_0}{\sqrt{\Omega^2 + c^2 Q^2}}}, \qquad \max\{|\Omega|, c Q\}\ll \tilde \Omega_0. \label{SMEq:DeffGammaeff}
\ee 
Plugging this ansatz back into Eq.~\eqref{SMEq:Deffeq} and following the calculations outlined in the previous section, we find that this equation is satisfied with logarithmic accuracy provided 
\be 
\tilde \Omega_0 \approx 0.818 \frac{\tilde\lambda^2 N N_0}{v_F^2} , \qquad c \approx 0.072v_F, \label{SMEq:tildeOmega0c}
\ee 
i.e., it is the same self-consistency condition as in Eq.~\eqref{SMEq:Omega0c}. This result is not surprising, since the relevant frequencies and momenta $(\omega, c \bq)$ in the integral are of the order of $\max\{|\Omega|,cQ\}$, leading to the mutual cancellation of large logarithmic factors from the equation. Next, plugging this ansatz into Eq.~\eqref{SMEq:Gammaeffeq}, we find with logarithmic accuracy
\be  
0 \approx \frac{\lambda}2 +    \frac{\tilde \lambda^3 N N_0}{8\pi^2 \Omega_0 c^2} \frac{v_F}{v_F + c} = \frac{\lambda}2 + 2.79\tilde \lambda \qquad \Longrightarrow \qquad \tilde \lambda \approx - \frac{\lambda}{5.57}. \label{SMEq:tildelambda}
\ee
Here we neglected the left-hand side of the equation, $\Gamma_{\text{eff}}(Q)$, due to its logarithmic smallness, and then used Eq.~\eqref{SMEq:tildeOmega0c}. The most relevant integration region in Eq.~\eqref{SMEq:Gammaeffeq} is $Q < q <\tilde \Omega_0/c$, which justifies our previous assumption. 

Logarithmic corrections in Eq.~\eqref{SMEq:DeffGammaeff} do not qualitatively change the expression for the fermion self-energy, Eq.~\eqref{SMEq:fermionSESC2D}. Hence, the Fermi-liquid behavior is preserved. 

We stress that while Eqs.~\eqref{SMEq:DeffGammaeff}-\eqref{SMEq:tildelambda} formally satisfy the system of self-consistent equations~\eqref{SMEq:Deffeq}-\eqref{SMEq:Gammaeffeq} at low energies, one should consider it as a guess rather than a mathematically rigorous solution.  Indeed, the perturbative correction~\eqref{SMEq:GammaTSC2D} increases the coupling as the energy scale decreases, while $\tilde \lambda$ changes sign compared to $\lambda$. This indicates that within our analysis, the solutions at the ultraviolet and infrared scales may not be smoothly connected, and the effective coupling may pass through the singularity at the scale $\sim\Omega_0$. Our analytical description completely fails at this scale. To make a smooth connection between the scales and reach the low-energy region, powerful numerical methods must be adopted, such as Quantum Monte Carlo. We leave this problem for future study.

\subsection{(d) The failure of the self-consistent approach for an isotropic boson}
Finally, we demonstrate that the same approach, with the correct power scaling~\eqref{SMEq:DTSC}, {\it does not} yield a self-consistent solution for a zero-momentum QCP when the boson is {\it isotropic}, i.e., {\it not projected} onto the transverse sector. The boson self-energy in this case is given by (we neglect vertex corrections for simplicity in this section)
\be  
\Pi(i\Omega, \bQ) - \Pi(0,0) = \lambda^2 N N_0 \int \frac{d\omega d^2 \bk}{(2\pi)^3}\left(\frac{|\omega + \Omega|}{\sqrt{(\omega+\Omega)^2 + v_F^2 (\bk + \bQ)^2}} - \frac{|\omega|}{\sqrt{\omega^2 + v_F^2 k^2}} \right)\frac{1}{\Omega_0^{1/2}(\omega^2 + c^2 k^2)^{3/4}},
\ee 
i.e., the transverse projector $P^T_{ij}$ is absent in comparison with Eq.~\eqref{SMEq:PiSC}. Consequently, we find that
\be 
\Pi(i\Omega,0) - \Pi(0,0) = \frac{\lambda^2 N N_0}{\Omega_0^{1/2} c^2} |\Omega|^{3/2} \frac{\Gamma(5/4)}{2 \pi^{7/2}\Gamma(3/4)} 2 F_{\Omega}\left( \frac{v_F}c  \right),
\ee
where $F_{\Omega}(x)$ is given by Eq.~\eqref{SMEq:FOmega}, and 
\be  
\Pi(0,Q) - \Pi(0,0) = \frac{\lambda^2 N N_0}{\Omega_0^{1/2}c^{1/2}}Q^{3/2}\frac{\Gamma(5/4)}{2\pi^{7/2}\Gamma(3/4)} F_{Q}\left(\frac{v_F}c\right), \qquad \text{with}
\ee 
\begin{align}  
F_{Q}(x) & = \int_0^1 \frac{dt}{t^{1/2} (1-t)^{1/4} (t x^2 + 1-t)^{1/4}} \int_0^{\infty}q dq \int_0^{2\pi} d\varphi \left\{\frac1{\left[q^2 + \frac{tx^2 (2q\cos \varphi+1)}{tx^2 + 1-t}   \right]^{1/4}}   - \frac1{q^{1/2}} \right\}  \label{SMEq:FQ} \\  &\approx -\frac{16 \pi}3\left\{ \begin{array}{cc} \displaystyle \frac{2 \pi^{1/2}\Gamma(5/4)}{3 \Gamma(3/4)}x - x^{3/2}, & x\ll1 \\[1.4em] \displaystyle \frac{\pi^{1/2}\Gamma(9/4)}{5 \Gamma(7/4)} \frac1{x}, & x\gg1 \end{array}     \right.   \approx \left\{ \begin{array}{cc} - 14.64 x + 16.76 x^{3/2}, & x\ll1 \\[1.4em] \displaystyle -\frac{7.32}{x}, & x\gg1. \end{array}     \right. \nonumber
\end{align}
The frequency dependence given by function $F_\Omega(x)$ is the same as before, apart from the factor of 2. The momentum dependence, $F_Q(x)$, is very different, since the factor $\cos^2\varphi$ under the integral is absent compared to Eq.~\eqref{SMEq:FQT}. As shown in Fig.~\ref{SMFig:F}(b) and in Eq.~\eqref{SMEq:FQ}, $F_Q(x) < 0$ for all $x$, making it impossible to find the self-consistent solution of the form of Eq.~\eqref{SMEq:DTSC}. This result is consistent with the perturbative analysis that uses bare Green's functions.

\end{widetext}

\end{document}